\documentclass[journal]{IEEEtran}
\def\subparagraph{} 
\usepackage{titlesec}
\titlespacing*{\section}{0pt}{*1}{*1}
\titlespacing{\subsection}{0pt}{*1}{*1}

\renewcommand{\thesubsubsection}{\arabic{subsubsection}}

\titleformat{\subsubsection}[runin]{\itshape}{\thesubsubsection)}{1em}{}
\titlespacing*{\subsubsection}{\parindent}{0pt}{*1}
\usepackage{verbatim}
\usepackage{amsmath,amssymb,amsfonts}

    \usepackage{verbatim}
    \usepackage{amssymb,amsfonts}
    \usepackage{amsthm}
    \usepackage[linesnumbered,ruled,vlined]{algorithm2e}

    \usepackage[frak=esstix]{mathalpha}
    \usepackage{algorithmic}
\SetAlFnt{\footnotesize}
\SetAlCapFnt{\small}
\usepackage{array}
\usepackage[caption=false,font=scriptsize,labelfont=sf,textfont=sf]{subfig}
\usepackage{textcomp}
\usepackage{stfloats}
\usepackage[sort&compress,numbers]{natbib}
\usepackage{acronym}
\usepackage[T1]{fontenc}
\usepackage{balance}
\usepackage{enumitem}
\usepackage{url}
\usepackage{cuted}
\usepackage[makeroom]{cancel}
\usepackage{lipsum}
\usepackage{verbatim}
\usepackage{graphicx}
\usepackage{cite}
\usepackage{amsmath,amssymb,amsfonts}
\usepackage{graphicx}
\usepackage{textcomp}
\usepackage{xcolor}
\usepackage{listings}
\usepackage{pdfpages}
\usepackage{mathrsfs}
\usepackage{bm}
\usepackage{booktabs}
\usepackage{multicol}
\usepackage{multirow}
\usepackage[caption=false]{subfig}
\usepackage{amsthm}
\theoremstyle{definition}
\usepackage{mathrsfs}
\def\BibTeX{{\rm B\kern-.05emb{\sc i\kern-.025em b}\kern-.08em
    T\kern-.1667em\lower.7ex\hbox{E}\kern-.125emX}}

\usepackage{tikz}
\usepackage{circuitikz}
\usetikzlibrary{positioning}
\usepackage{pgfkeys}
\usetikzlibrary{spy}

\usepackage{pgfplots}

\usepgfplotslibrary{groupplots} 
\usepgfplotslibrary[groupplots] 
\usetikzlibrary{pgfplots.groupplots} 
\usetikzlibrary[pgfplots.groupplots] 

\pgfplotsset{compat=newest}
\pgfplotsset{plot coordinates/math parser=false}
\usepackage{pgfplotstable}
\newlength\figureheight
\newlength\figurewidth

\usetikzlibrary{positioning}
\usetikzlibrary{patterns}

\usetikzlibrary{arrows.meta,positioning}
\tikzset{block/.style={draw, rectangle, fill=cyan!90,
        minimum height=2em, minimum width=3em},
    sum/.style={draw, circle, node distance=1cm},
    input/.style={coordinate},
    output/.style={coordinate},
    pinstyle/.style={pin edge={to-,thin,black}},
        saturation block/.style={%
            draw,
            path picture={
                \pgfpointdiff{\pgfpointanchor{path picture bounding box}{south west}}%
                {\pgfpointanchor{path picture bounding box}{north east}}
                \pgfgetlastxy\x\y
                \tikzset{x=\x*.4, y=\y*.4}
                \draw [very thin] (-1,0) -- (1,0) (0,-1) -- (0,1);
                \draw [very thick] (-1,-.7) -- (-.7,-.7) -- (.7,.7) -- (1,.7);
            },
        }
    }

\tikzset{%
        rateLimit block/.style={%
            draw,
            path picture={
                \pgfpointdiff{\pgfpointanchor{path picture bounding box}{south west}}%
                {\pgfpointanchor{path picture bounding box}{north east}}
                \pgfgetlastxy\x\y
                \tikzset{x=\x*.4, y=\y*.4}
                \draw [very thin] (-1,0) -- (1,0) (0,-1) -- (0,1);
                \draw [very thick] (-1,-1) -- (1, 1);
            },
        }
    }

     \usetikzlibrary{svg.path}
     \usepackage{scalerel}
\usepackage{hyperref}
    \definecolor{orcidlogocol}{HTML}{A6CE39}
    \tikzset{
      orcidlogo/.pic={
        \fill[orcidlogocol] svg{M256,128c0,70.7-57.3,128-128,128C57.3,256,0,198.7,0,128C0,57.3,57.3,0,128,0C198.7,0,256,57.3,256,128z};
        \fill[white] svg{M86.3,186.2H70.9V79.1h15.4v48.4V186.2z}
                     svg{M108.9,79.1h41.6c39.6,0,57,28.3,57,53.6c0,27.5-21.5,53.6-56.8,53.6h-41.8V79.1z M124.3,172.4h24.5c34.9,0,42.9-26.5,42.9-39.7c0-21.5-13.7-39.7-43.7-39.7h-23.7V172.4z}
                     svg{M88.7,56.8c0,5.5-4.5,10.1-10.1,10.1c-5.6,0-10.1-4.6-10.1-10.1c0-5.6,4.5-10.1,10.1-10.1C84.2,46.7,88.7,51.3,88.7,56.8z};
      }
    }

    \newcommand\orcidicon[1]{\href{https://orcid.org/#1}{\mbox{\scalerel*{
    \begin{tikzpicture}[yscale=-1,transform shape]
    \pic{orcidlogo};
    \end{tikzpicture}
    }{|}}}}

\begin{document}

\title{Demystifying 5G Polar and LDPC Codes : \\A Comprehensive Review and Foundations}

\author{\IEEEauthorblockN{ Mody~Sy~{\orcidicon{0000-0003-2841-2181}}}

\thanks{Manuscript prepared on February 11, 2025. The author is with the Department of Communication Systems, EURECOM, BIOT, 06410, France (e-mails: mody.sy@eurecom.fr).}
}
\makeatletter
\patchcmd{\@maketitle}
  {\addvspace{0.5\baselineskip}\egroup}
  {\addvspace{-1\baselineskip}\egroup}
  {}
  {}
\makeatother

\maketitle

\begin{abstract}
Understanding how 5G networks correct errors is no trivial matter. At the heart of the process lie two sophisticated families of codes: LDPC and polar codes. This paper opens the black box, not only by explaining how these codes are designed, but also by showing how they are encoded and decoded in practice. To describe the research covered in this review, we present a detailed survey of the literature supplemented with insights that are often buried deep within technical standards. These foundations are not just historical footnotes: they remain relevant to research on error correction for future wireless systems. In bringing clarity to these building blocks, we aim to help engineers and researchers navigate what is both a complex and increasingly vital part of wireless communication.
\end{abstract}


\section{Introduction}
\label{sec:introduction}

\IEEEPARstart{C}{hannel} coding was born the day Shannon showed that even noisy communication could be made nearly error-free provided that suitable codes and sufficiently long blocks are used. His 1948 paper laid the foundation~\citep{Shannon48}: by embedding redundancy into transmitted data, one could correct errors as long as the data rate stayed below a theoretical threshold, now known as the channel capacity.
Since then, the quest has been clear: how do we push transmission rates as close as possible to that limit while still taming noise, distortion, and interference? Two tools remain essential: modulation, which shapes the physical signal, and channel coding, which reshapes the message itself.
Many practical channel codes are linear and can be encoded systematically, so that the information bits appear explicitly in the codeword. Among the many varieties, Reed--Solomon~\citep{Whitaker1991}, Hamming~\citep{Hamming1950}, Bose--Chaudhuri--Hocquenghem (BCH) \citep{Hocquenghem1969}, \acs{CRC}, and \acs{LDPC} \citep{Gallager63} codes have emerged as cornerstones, each bringing unique properties suited to different applications. These codes can be expressed algebraically, via matrices, or visually, as factor graphs.

In practice, the matrix-based representation is most common, relying on two fundamental constructs: the \textit{generator matrix} and the \textit{parity-check matrix}.

A linear block code $\mathcal{C}$ of length $N$
and dimension $K$ over the binary field $\mathbb{F}_2$
is defined as the image of a generator matrix
$\mathbf{G}\in\mathbb{F}_2^{K\times N}$ of rank $K$:
\begin{equation}
\mathcal{C}=
\left\{
\mathbf{c}=\boldsymbol{m}\mathbf{G}
\mid \boldsymbol{m}\in\mathbb{F}_2^K
\right\}.
\end{equation}

A generator matrix $\mathbf{G}$ is said to be in
\emph{systematic form} if it can be expressed as
$\mathbf{G}=[\mathbf{I}_K\;\;\mathbf{P}]$,
where $\mathbf{I}_K$ denotes the $K\times K$
identity matrix and
$\mathbf{P}\in\mathbb{F}_2^{K\times(N-K)}$
is the associated parity matrix.

More generally, a generator matrix need not be
systematic. After a suitable permutation of its
columns, any full-row-rank generator matrix can
be written as
$\mathbf{G}=[\mathbf{P}'\;\;\mathbf{M}]$,
where $\mathbf{M}\in\mathbb{F}_2^{K\times K}$
is invertible and
$\mathbf{P}'\in\mathbb{F}_2^{K\times(N-K)}$.
The column permutation corresponds to the same
permutation of the codeword coordinates.

The code $\mathcal{C}$ can alternatively be specified
by a parity-check matrix
$\mathbf{H}\in\mathbb{F}_2^{(N-K)\times N}$
of rank $N-K$, satisfying
\begin{equation}
\mathcal{C}=
\left\{
\mathbf{c}\in\mathbb{F}_2^N
\mid \mathbf{H}\mathbf{c}^{\mathsf T}=\mathbf{0}
\right\}.
\end{equation}

When $\mathbf{G}$ is in systematic form, a
corresponding parity-check matrix is
$\mathbf{H}=[\mathbf{P}^{\mathsf T}\;\;
\mathbf{I}_{N-K}]$,
ensuring the orthogonality condition
$\mathbf{G}\mathbf{H}^{\mathsf T}=\mathbf{0}$.

Given a hard-decision received vector
$\boldsymbol{r}\in\mathbb{F}_2^N$,
its \emph{syndrome} is computed as
$\boldsymbol{s}=\mathbf{H}\boldsymbol{r}^{\mathsf T}$.
If $\boldsymbol{s}=\mathbf{0}$, then
$\boldsymbol{r}$ is a valid codeword, although an
error that maps one codeword to another remains
undetected. Otherwise, $\boldsymbol{s}\neq\mathbf{0}$
indicates the presence of errors, and the syndrome
can be used to help infer the error pattern.
\begin{table}[htbp]
  \centering
  \caption{Channel coding schemes for {\em transport channels} (TrCHs)\citep{3GPP38212}.}
  \begin{tabular}{l|l}
    \toprule
    \textbf{TrCH} & \textbf{Coding scheme} \\
    \hline
    \acs{UL-SCH} & LDPC \\
    \acs{DL-SCH} & LDPC \\
    \acs{PCH} & LDPC \\
    \hline
    \acs{BCH} & Polar code \\
    \bottomrule
  \end{tabular}%
  \label{tab:5.3-1}%
\end{table}
In 5G New Radio (NR), these principles take on practical form. Information, whether data or control, is encoded at the physical layer (\acs{PHY}) before transmission; transport blocks are delivered to the PHY by the \acs{MAC}, enabling robust communication over unpredictable wireless channels. The coding scheme used in 5G is a carefully engineered blend of error detection, correction, rate matching, interleaving, and mapping onto physical resources.
Tables \ref{tab:5.3-1} and \ref{tab:5.3-2} summarize the coding choices made in the 5G NR standard for various transport and control channels \citep{3GPP38212}.
\begin{table}[htbp]
  \centering
  \caption{Usage of channel coding schemes within  5G NR UL/DL control channels \citep{3GPP38212}.}
    \begin{tabular}{l|l}
    \toprule
    \textbf{Control Information} & \textbf{Coding scheme }\\
    \hline
    \acs{DCI} & Polar code \\
    \hline
    \acs{UCI} & Polar or small-block coding \\
    \bottomrule
    \end{tabular}%
  \label{tab:5.3-2}%
\end{table}%
This paper serves as a guide and reference for both practitioners and scholars interested in understanding channel coding and decoding schemes, particularly the Low-Density Parity-Check (LDPC) and polar coding methods adopted in \acs{5G} and beyond, in line with 3GPP standard specifications. It highlights important information that can be difficult to extract from the {\em technical specification} documents established by the standard.

Notably, we enrich the foundational reviews of the main features of these two coding schemes with many standard-specific details, such as rate adaptation procedures, the application of Cyclic Redundancy Check (CRC), and aspects like decoding algorithms. The refinement and deeper understanding of these coding schemes are particularly relevant to research on error correction in future wireless systems.

The standard-specific descriptions refer to TS 38.212 V16.2.0 (Release 16), and the literature discussed here includes publications through 2023.

The article is structured as follows. Sections II and III lay out the foundations and detailed reviews of 5G NR LDPC codes and 5G NR Polar codes respectively, and Section IV concludes the paper.

Furthermore, the following is a list of acronyms that the reader will encounter throughout the manuscript:
\setlength{\arrayrulewidth}{0.4pt} 

\begin{footnotesize}
\begin{center}
\begin{acronym}

\acro{5G}{{Fifth Generation of Wireless Cellular Technology}}
\acro{NR}{{New Radio}}
\acro{BCH}{{Broadcast Channel}}
\acro{BLER}{{Block Error Rate}}
\acro{BP}{{Belief Propagation}}
\acro{CA-SCL}{{CRC-aided Successive Cancellation List}}
\acro{CRC}{{Cyclic Redundancy Check}}
\acro{DCI}{{Downlink Control Information}}
\acro{DL-SCH}{{Downlink Shared Channel}}
\acro{LBP}{{Layered Belief Propagation}}
\acro{LDPC}{{Low-Density Parity-Check}}
\acro{LLR}{{Log-Likelihood Ratio}}
\acro{MAC}{{Medium Access Control}}

\acro{OFDM}{{Orthogonal Frequency Division Multiplexing}}
\acro{PCH}{{Paging Channel}}
\acro{PHY}{{Physical Layer}}
\acro{SC}{{Successive Cancellation}}
\acro{SCL}{{Successive Cancellation List}}
\acro{TB}{{Transport Block}}
\acro{UCI}{{Uplink Control Information}}
\acro{UL-SCH}{{Uplink Shared Channel}}
\acro{URLLC}{{Ultra-Reliable-Low-Latency Communication}}

\end{acronym}
\end{center}
\end{footnotesize}

\section{5G NR LDPC Codes}
\acs{LDPC} codes received limited attention for several decades after their introduction. LDPC codes were originally proposed by Gallager in the early 1960s~\citep{Gallager63}. At that time, the codes might have been overlooked because contemporary investigations in concatenated coding overshadowed LDPC codes and because the hardware of the time could not support effective decoder implementations\citep{Moon2005}. They therefore received limited attention until the 1990s, when interest in iterative decoding was renewed by turbo codes\citep{Berrou1993}.
Since then, {LDPC} codes have shown interesting performance and a relatively uncomplicated implementation.
MacKay, working on Turbo codes at that time, renewed interest in {LDPC} codes \citep{MacKay1997} and brought {LDPC} codes back into focus. This article by MacKay presents constructions of {LDPC} codes and shows their good performance.
Later, Luby and colleagues introduced irregular {LDPC} codes \citep{Luby1998} characterized by a parity-check matrix for which the distribution of the number of non-zero elements per row and/or column is not uniform. {LDPC} codes are linear block codes based on sparse parity-check matrices. In recent years, {LDPC} codes have attracted more attention because of their efficient decoding algorithms, excellent error-correcting capability, and their performance close to the Shannon limit for large code lengths.
{LDPC} coding  is currently adopted in 5G NR for both uplink and downlink shared transport channels. Given that 5G must support high data rates, a wide range of block sizes and code rates for data channels, and {\em hybrid automatic repeat request} (HARQ), {LDPC} codes provide a suitable coding framework for these requirements. Indeed, the base graphs defined in 3GPP TS 38.212 \citep{3GPP38212} define structured parity-check matrices, which support HARQ, a wide range of information block lengths, and rate-compatible transmission. While  Polar codes are applied to 5G {NR} control channels, {LDPC} codes are suitable for 5G NR shared channels due to their high throughput, low latency, low decoding complexity and rate compatibility. Their suitability for \acs{URLLC} depends on the selected block length, code rate, decoder, and operating conditions; a low error floor cannot be guaranteed for every configuration.
\subsection{State-of-the-Art LDPC Codes}
In reviewing the literature, significant efforts have been directed towards enhancing the error correction performance of 5G communication systems.
The discourse surrounding channel coding in cellular systems, particularly for 5G, initially emphasized turbo, \acs{LDPC}, and polar codes, with LDPC codes adopted for eMBB data and polar codes for control \citep{Shao2019}.  In this regard, \citet{Srisupha2022} developed an experimental kit demonstrating {LDPC} encoding processes, emphasizing flexibility between software and hardware {LDPC} encoders. Subsequently, \citet{Belhadj2021} compared error correction performance between {LDPC} and polar codes in 5G {\em machine-to-machine  ({M2M}) communications}, highlighting specific requirements for different {M2M} applications. Later on, \citet{Li2023} proposed dynamic scheduling strategies to reduce decoding complexity and improve error correction performance for short {LDPC} codes. Next, from the perspective of accelerated {LDPC} decoding, \citet{Tian2022} presented a base graph-based static scheduling method for layered decoding of 5G {LDPC} codes, achieving notable reductions in iteration numbers and performance enhancements. In \citep{Hamidi-Sepehr2018, bae2019}, authors investigated the structure and features of the base graphs \citep{3GPP38212}, showing that the usage of a circularly shifted identity matrix known as the permutation matrix can greatly reduce the memory requirement for implementation. The sparsity pattern of the 5G {LDPC} base graphs permits some check rows with disjoint variable-node neighborhoods to be grouped for parallel layered processing. This property can reduce decoding latency, but it does not imply that all base-graph rows are mutually orthogonal. For both base graphs, the first two circulant columns are omitted from the LDPC encoder output. This puncturing is part of the NR code design and must be considered together with code construction and rate matching; puncturing alone does not guarantee improved error-correction performance~\citep{bae2019}. Additionally, 5G {LDPC} fully utilizes the double diagonal structure of the base graphs. Due to the characteristic of base graphs, the double-diagonal structure can make {LDPC} encoding more efficient. Additionally, \citet{Nguyen2019} proposed a novel efficient encoding method and a high-throughput low-complexity encoder architecture for 5G NR \acs{LDPC}.

Furthermore, efforts have been directed towards optimizing implementations of these codes for heightened performance on software and hardware targets. For instance, \citet{Liao2021} present a high-throughput LDPC encoding on a single {\em graphics processing unit} (GPU), while \citet{Tarver2021} explore GPU-based LDPC decoding, showcasing potential for high throughput and low latency applications in 5G and beyond. Additionally, hardware architectures have been developed to efficiently decode LDPC codes, as seen in the work of \citet{Nadal2021}, who propose a highly parallel {\em field-programmable gate array} (FPGA) architecture. Conversely, \citet{Xu2019} focus on software decoding with SIMD acceleration on Intel Xeon CPUs, achieving notable throughput with low latency. \citet{sy2023_ldpc} proposed optimization strategies for low-latency 5G LDPC decoding over GPPs. Meanwhile, \citet{Li2021_ldpc} addressed the challenge of achieving high throughput rates, proposing a multicore LDPC decoder architecture achieving up to 1 Tb/s throughput for beyond 5G systems, while \citet{Aronov2019} compare LDPC decoding performance between GPU and FPGA platforms, stressing the need for further optimization, particularly in reducing latency for GPU-based solutions.

Moreover, the selection of coding schemes for 5G eMBB services has been a focus of interest, with {\em quasi-cyclic} LDPC and polar codes chosen for data and control channels, respectively. \citet{Rao2019} highlighted the importance of these codes, particularly QC-LDPC, and \citet{Wu2019} proposed an efficient QC-LDPC implementation for 5G NR, enhancing throughput by matrix pruning. \citet{Ivanov2023} introduced a novel concatenated code construction comprising outer and inner LDPC codes, demonstrating reduced decoding complexity and superior performance. Additionally, \citet{Song2023} emphasized the significance of well-designed LDPC codes, particularly QC-LDPC codes, in approaching channel capacity and enabling high-speed data transmission. In parallel, LDPC codes' adoption in 5G standards underscores their importance in broadcasting and cellular communication systems  \citet{Ahn2019}.  Moreover, \citet{Cui2021} tackled the challenge of designing high-performance and area-efficient decoders, while \citet{Trung2019} and \citet{Jayawickrama2022} proposed adaptations and improved algorithms for LDPC decoding in 5G NR. Additionally, \citet{Li2020} focused on LDPC code design for specific 5G scenarios, proposing optimization methods and an improved decoding algorithm. \citet{Wu02019} provided insights into LDPC decoding latency in 5G NR, guiding decoder design to meet high throughput and low latency requirements. \citet{Sun2018} proposed a hybrid decoding algorithm for LDPC codes in 5G, where {normalized min-sum algorithm} (NMSA) decoding and linear approximation are combined, with only a slight increase in complexity for NMSA and an improved performance much closer to {\em belief propagation} \acs{BP} decoding, especially for low-rate codes.

Recently, there has been a growing interest in applying deep learning techniques to various aspects of 5G communications, including the protograph LDPC decoding methods studied by \citet{Dai2021}. \citet{Shah2021} proposed {\em neural layered min-sum} (NLMS) algorithms to enhance the decoding performance of 5G LDPC codes, leveraging {\em deep neural networks} (DNNs) to learn normalization and offset parameters. \citet{Tang2021} introduced a scheme combining model-driven deep learning with traditional BP decoding algorithms to adapt LDPC codes for different 5G scenarios. Lastly, \citet{Andreev2021} investigated the application of DNNs to improve the decoding algorithms of short
QC-LDPC codes in the 5G standard, addressing the curse of dimensionality problem and enhancing performance.
\subsection{Foundations and Fundamentals}
The LDPC code is presented in matrix form
$[\mathbf{G}, \mathbf{H}]$, where $\mathbf{G}$ is the generator matrix of the code and $\mathbf{H}$ is the {\em parity-check matrix} (PCM). The parity-check matrix $\mathbf{H}$ is sparse, containing very few ones. It can be represented in the form of a {\em Tanner graph}. This graph consists of two types of nodes: {\em bit nodes} (BNs) and {\em check nodes} (CNs) which are connected by edges. The BNs and CNs correspond, respectively, to the columns (codeword bits) and rows (parity constraints) of the matrix $\mathbf{H}$. A variable node $i$ is connected to a check node $j$ if $H_{j,i}=1$. This graph is bipartite, since nodes of the same type cannot be connected (i.e., a CN cannot be connected to another CN). Tanner graphs are commonly used to represent the parity matrix in LDPC codes. A Tanner graph leads to decoding algorithms of fairly low complexity \citep{Moon2005}.

\begin{equation}\label{eqn:pcm_ldpc}
\mathbf H=
\left[\begin{array}{llllllll}
0 & 1 & 0 & 1 & 1 & 0 & 0 & 1 \\
1 & 1 & 1 & 0 & 0 & 1 & 0 & 0 \\
0 & 0 & 1 & 0 & 0 & 1 & 1 & 1 \\
1 & 0 & 0 & 1 & 1 & 0 & 1 & 0
\end{array}\right]
\end{equation}

The number of variable nodes equals the full codeword length ${N}$, that is, the number of columns of $\mathbf H$, including punctured positions. The number of check nodes equals its number of rows $M$, and the code dimension is ${K}={N}-\operatorname{rank}_{\mathbb F_2}(\mathbf H)$. Only when the rows are independent does $M={N}-{K}$.
For the matrix in (\ref{eqn:pcm_ldpc}), the rank over $\mathbb F_2$ is 3, so the four check nodes define an $(8,5)$ code.
\begin{figure}[ht]
    \centering
      \includegraphics[width=0.8\linewidth]{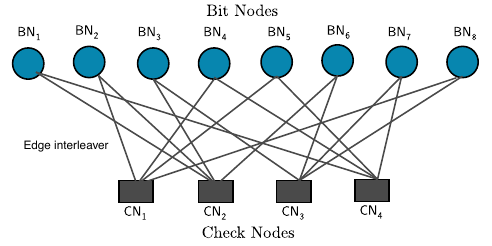}
       \caption{Tanner graph corresponding to the $4\times8$ parity-check matrix $\mathbf H$.}
       \label{fig:tanner_graph}
\end{figure}

The most important element in the realization of a good LDPC code is the  $\mathbf H$ matrix which will condition the quality of the iterative decoding.
There are four important parameters to respect in order to obtain a good $\mathbf H$ parity-check matrix.
\begin{enumerate}
  \item \emph{Code Rate}.
  The code rate is ${R}=1-\operatorname{rank}_{\mathbb F_2}(\mathbf H)/{N}$, where ${N}$ is the number of columns. Adding nonzero entries alone does not necessarily increase the rate.
  For a regular parity-check matrix with column weight $w_c$, row weight $w_r$, and $M$ rows, edge counting gives $\mathrm Nw_c=Mw_r$. Thus, the design rate is $1-w_c/w_r$, and it equals the actual rate when the rows of $\mathbf H$ are linearly independent.
  \item \emph{short cycles (girth)}. The generation of short cycles in the equivalent Tanner graph of the code must be avoided. A cycle is a closed path alternating between variable nodes and check nodes, with no repeated vertex except the starting and ending vertex.
We call girth the minimum cycle length that can be encountered in a Tanner graph.
Short cycles are very penalizing because they involve few intermediate nodes, and thus the extrinsic information they generate during decoding becomes quickly and strongly correlated.
\item \emph{Size of $\mathbf H$}. Increasing the block length can improve performance for well-designed code families, but matrix size alone determines neither the code rate nor the decoding performance.
\item \emph{Construction algorithm of $\mathbf H$}.
Parity-check matrices can be constructed using random, graph-based, or structured methods. Gallager constructions~\citep{Gallager63} use random permutations, whereas the {\em progressive edge-growth} (PEG) method~\citep{Hu2005PEG} adds edges progressively to favor large local girth. PEG can construct both regular and irregular Tanner graphs. Quasi-cyclic constructions use arrays of circulant matrices, often circular permutations of the identity matrix~\citep{Wu2019, Zhang2010}.
QC-LDPC codes belong to the class of structured codes that are relatively easier to implement without significantly compromising the performance of the code. Well-designed QC-LDPC codes can provide strong bit-error and block-error performance, but their performance and error floors depend on the construction and decoding configuration. These codes also offer merits in decoder hardware implementation due to their cyclic symmetry, which results in simple regular interconnection and modular structure\citep{Ahmadi2018}.
\end{enumerate}

\subsection{3GPP 5G LDPC Codes}
NR LDPC code is a family of QC-LDPC codes. It is constructed from a matrix named $\mathbf H_{\mathsf{BG}}$ of dimension $M \times N$ called {\em base graph} matrix $\mathbf{BG}$. The base graph is selected according to the target code rate and transport-block payload size; the lifting size is then selected for the segmented code blocks. Thus, for
$\mathbf{BG}_\mathsf 1$ ($N = 68$, $M = 46$) and for $\mathbf{BG}_\mathsf 2$ ($N = 52$, $M = 42$). $\mathbf{BG}_1$ supports encoder input lengths up to $8448$ bits, whereas $\mathbf{BG}_2$ supports lengths up to $3840$ bits. The exact selection depends on the transport-block size and target code rate, as specified below; rate matching permits transmitted rates beyond the nominal mother-code rates.

For $\mathbf{BG}_\mathsf{1}$, ${K}=22Z_c$ and for $\mathbf{BG}_\mathsf{2}$, ${K}=10Z_c$, where ${K}$ is the encoder input length, including code-block CRC and filler bits, and $Z_c$ is the lifting size shown in Table \ref{tab:lift_size}. There are 51 lifting sizes from 2 to 384 for each base graph. Both $\mathbf{BG}_\mathsf{1}$ and $\mathbf{BG}_\mathsf{2}$ have the same block structure. The columns include information columns, core parity columns, and extension parity columns. The rows are divided into core check rows and extension check rows.

\begin{figure}[ht]
    \centering
         \includegraphics[width=0.8\linewidth]{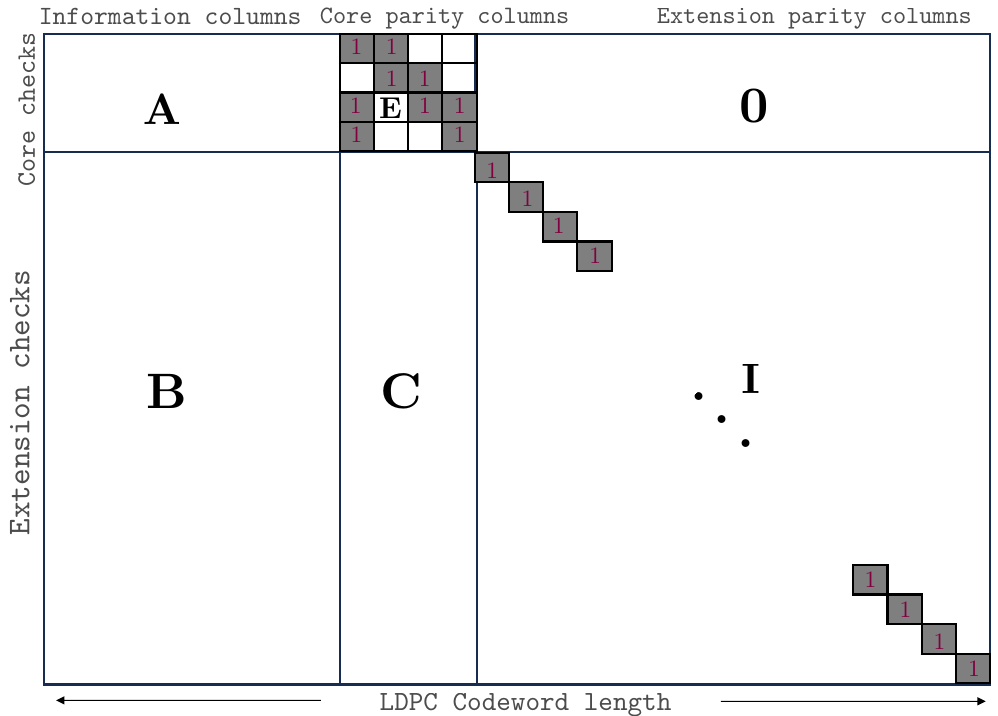}
         \caption{ 3GPP NR Base graphs structure.}
       \label{fig:BG_struct}
\end{figure}

For $\mathbf{BG}_\mathsf{1}$, $\mathbf A$ is a $4\times 22$ matrix; $\mathbf E$ is a $4 \times 4$ matrix; $\mathbf O$ is a $4 \times 42$ all-zero matrix; $\mathbf B$ is a $42 \times 22$ matrix; $\mathbf C$ is a $42 \times 4$ matrix; $\mathbf I$ is a $42 \times 42$ identity matrix;\\
For $\mathbf{BG}_\mathsf{2}$, $\mathbf A$ is a $4 \times 10$ matrix; $\mathbf E$ is a $4\times 4$ matrix; $\mathbf O$ is a $4 \times 38$ all-zero matrix;
$\mathbf B$ is a $38 \times10$ matrix; $\mathbf C$ is a $38 \times 4$ matrix; $\mathbf I$ is a $38\times38$ identity matrix.
Submatrix $\mathbf E$ is the structured core-parity submatrix that facilitates encoding. The common block structure of the NR LDPC base graphs is shown in Figure~\ref{fig:BG_struct}. The standard distinguishes the binary base-graph connectivity from the tabulated shift coefficients. In the shift-coefficient representation used here, $-1$ denotes an absent edge, whereas a nonnegative entry represents a circulant shift coefficient.
In 3GPP TS 38.212~\citep{3GPP38212}, the set of allowed lifting sizes for each index of iLS is shown in Table~\ref{tab:lift_size}. A missing edge is denoted by $-1$. For each nonnegative tabulated shift coefficient $V_{i,j}$, the actual circulant shift is $p_{i,j}=V_{i,j}\bmod Z_c$, with $0\leq p_{i,j}<Z_c$.

\begin{table}[htbp]
  \centering
  \caption{Sets of LDPC lifting sizes \citep{3GPP38212}.}
    \begin{tabular}{c l}
    \toprule
    Set index (iLS) & Set of lifting sizes (Z) \\
    \hline
    0     & 2, 4, 8, 16, 32, 64, 128, 256 \\
    1     & 3, 6, 12, 24, 48, 96, 192, 384 \\
    2     & 5, 10, 20, 40, 80, 160, 320 \\
    3     & 7, 14, 28, 56, 112, 224 \\
    4     & 9, 18, 36, 72, 144, 288 \\
    5     & 11, 22, 44, 88, 176, 352 \\
    6     & 13, 26, 52, 104, 208 \\
    7     & 15, 30, 60, 120, 240 \\
    \bottomrule
    \end{tabular}%
  \label{tab:lift_size}
\end{table}%
The smaller base graph of $\mathbf{BG}_2$ can reduce processing requirements for suitable block lengths and rates. However, decoding latency also depends on the lifting size, scheduling, iteration count, quantization, and architecture; the base-graph dimensions alone do not guarantee lower latency~\citep{Ahmadi2018}.\\
In this paragraph, $\mathbf H_{\mathsf{BG}}$ denotes the shift-coefficient representation, rather than the binary connectivity matrix. Each nonnegative entry $V_{i,j}$ is reduced modulo $Z_c$ before lifting. The parity-check matrix $\mathbf H$ is obtained according to the following rules.
\begin{itemize}
  \item Each element of value $-1$ in $\mathbf H_{\mathsf{BG}}$ is replaced by an all-zero matrix of size $Z_c \times  Z_c$.
  \item Each nonnegative entry whose reduced shift is $p_{i,j}=0$ is replaced by an identity matrix  $\mathbf I$  of size  $Z_c \times  Z_c$.
  \item Each nonnegative entry whose reduced shift is $1\leq p_{i,j}<Z_c$ is replaced by a circular permutation matrix $\mathbf I(\mathsf{p_{i,j}})$ of size  $Z_c \times  Z_c$,
  where $i$ and $j$ are the row and column indices of the element, and $\mathbf I(\mathsf{p_{i,j}})$ is obtained by circularly shifting the identity matrix $\mathbf I$ of size  $Z_c \times  Z_c$ to the right $\mathsf{p_{i,j}}$ times \citep{3GPP38212}.
\end{itemize}

\begin{figure*} [!ht]
    \centering
  \subfloat[\scriptsize {Transmitter end.}\label{fig:tx_ldpc}]{ 
  \includegraphics[width=0.8\linewidth]{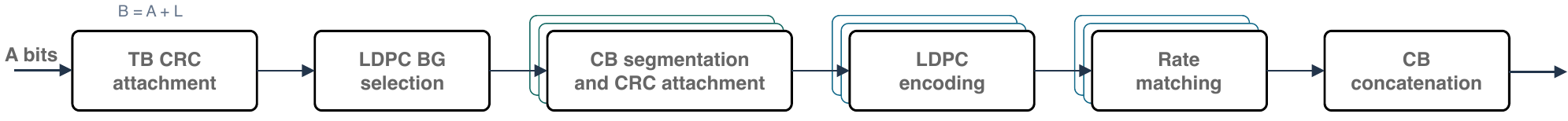}
  }
 \\
  \subfloat[\scriptsize {Receiver end.}\label{fig:rx_ldpc}]{  
    \includegraphics[width=0.8\linewidth]{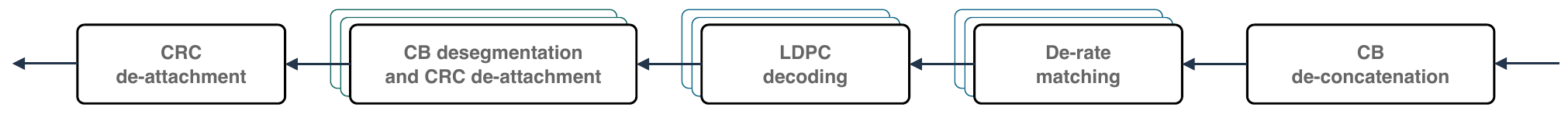}
    }
\caption{Conceptual illustration of 5G LDPC transceiver chain.}
\label{fig:ldpc_tranceiver_chain}
\end{figure*}

The main advantage of using a circularly shifted identity matrix is that it can reduce the memory requirement for implementation~\citep{bae2019,wang2021}.\\
To simplify, a small example can be used to explain the principle of obtaining the parity-check matrix $\mathbf H$. Hence,  assuming that $\mathbf H_\mathsf{{BG}}$ is a given base graph matrix with lifting size $Z_c$=4,
\begin{equation}
\mathbf H_{\mathsf{BG}}=\left[\begin{array}{cccccc}
\textcolor{cyan}{2} & \textcolor{violet}{-1} & \textcolor{orange}{1} & \textcolor{brown}{3} & \textcolor{green}{0} & \textcolor{violet}{-1} \\
\textcolor{orange}{1} & \textcolor{green}{0} & \textcolor{violet}{-1} & \textcolor{green}{0} & \textcolor{green}{0} & \textcolor{green}{0} \\
\textcolor{violet}{-1} & \textcolor{brown}{3} & \textcolor{cyan}{2} & \textcolor{orange}{1} & \textcolor{violet}{-1} & \textcolor{green}{0}
\end{array}\right],
\end{equation}
the corresponding parity-check matrix $\mathbf H$ is:
\begin{equation}
\includegraphics[width=0.8\linewidth]{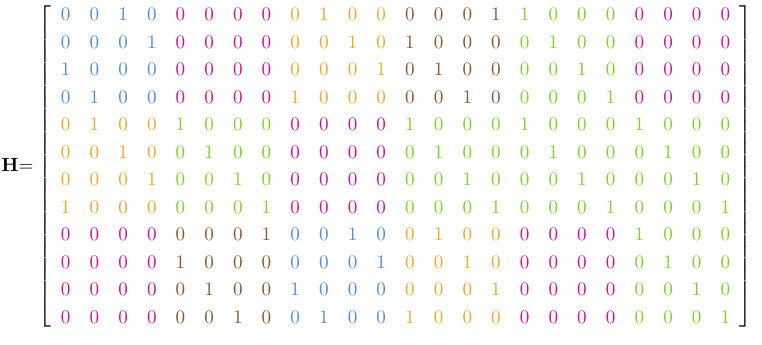}.
\end{equation}
The LDPC transmission chain for MAC/PHY processing is depicted in Figure~\ref{fig:ldpc_tranceiver_chain}. The transmitter for PUSCH/PDSCH includes transport-block CRC attachment, LDPC base-graph selection, code-block segmentation and CRC attachment, LDPC encoding, rate matching, and code-block concatenation. The receiver performs the corresponding LLR recovery and decoding operations.
At the transmitter end, we can list the following components.
\begin{enumerate}
  \item {\em Transport block CRC attachment}.\\
\acs{CRC} is an error-detection code used to detect residual errors after decoding. The entire transport block is used to calculate the CRC parity bits. Assume that the transport message before CRC attachment is $a(0), a(1),\ldots, a(A-1)$, where $A$ is the size of the transport block message. Parity bits are  $p(0), p(1),\ldots, p(L-1)$, where $L$ is the number of parity bits. The parity bits are generated by one of
the following cyclic generator polynomials\citep{3GPP38212}.
If $A > 3824$, the generator polynomial $\mathrm g_\mathsf{CRC24A}(\beta)$ is used.
\begin{equation}
\begin{aligned}
\mathrm g_{\mathsf{CRC24A}}(\beta)={}&\beta^{24}+\beta^{23}+\beta^{18}+\beta^{17}+\beta^{14}\\
&+\beta^{11}+\beta^{10}+\beta^7+\beta^6+\beta^5\\
&+\beta^4+\beta^3+\beta+1.
\end{aligned}
\end{equation}
The CRC length is $L=24$.
Otherwise, the generator polynomial $\mathrm g_{\mathsf{CRC16}}(\beta)$ is used.
\begin{equation}
\displaystyle\mathrm g_\mathsf{CRC16}(\beta)=\left[\beta^{16}+\beta^{12}+\beta^5+1\right].
\end{equation}
The CRC length is $L=16$.
The message bits after CRC attachment are $b(0),b(1),\ldots,b(B-1)$, where $B=A+L$.
\begin{equation}
b_k=\left\{\begin{array}{ll}
a_k, & k=0,1, \ldots, A-1 \\
p_{k-A}, & k=A, A+1, \ldots, A+L-1.
\end{array}\right.
\end{equation}
The CRC detects residual errors after LDPC decoding. Its undetected-error probability depends on the CRC polynomial and the residual error patterns; it should not be identified with the BLER.
\item {\em LDPC base graph selection}.\\
LDPC $\mathbf{BG}$ is selected based on the transport block message size $A$ and transport block coding rate ${R}$. If $A\leq292$, or if $A\leq3824$ and ${R}\leq0.67$, or if ${R} \leq 0.25$, LDPC $\mathbf{BG}_\mathsf{2}$ is used. Otherwise, LDPC $\mathbf{BG}_\mathsf{1}$ is used \citep{3GPP38212}.
\item Code block segmentation and code block CRC attachment.\\
The input message to CB segmentation is a transport message with CRC, denoted as $b(0), b(1),\ldots, b(B-1)$, where $B$ is the input message length. Assume that the maximum code block length is $\mathsf K_{cb}$, where $\mathsf K_{cb} = 8448$ for $\mathbf{BG}_\mathsf{1}$ and $\mathsf K_{cb} = 3840$ for $\mathbf{BG}_\mathsf{2}$. Code block segmentation is based on the following rules. \\
Assume that $C$ is the number of code blocks, $L$ denotes the code-block CRC length, and $B^\prime$ is total number of bits after code-block CRC attachment:\\
\texttt{if} $B\leq\mathsf K_{cb}$,
\begin{equation}
C=1,\qquad L=0,\qquad B'=B.
\end{equation}
\texttt{Otherwise,}
\begin{equation}
L=24,\qquad
C=\left\lceil\frac{B}{\mathsf K_{cb}-L}\right\rceil,
\qquad B'=B+CL.
\end{equation}

Each block contains ${K}'=B'/C$ information and CRC bits before filler insertion.

The output of code block segmentation is
$c_r(0),c_r(1),\ldots,c_r(\mathrm K_r-1)$,
where $\mathrm K_r={K}$ is the encoder input length,
including filler bits.

The lifting size $Z_c$ is the smallest value in
Table~\ref{tab:lift_size} satisfying
$\mathsf K_b Z_c\geq{K}'$.
The parameter $\mathsf K_b$ is given in
Table~\ref{tab:kb_value}.

For $\mathbf{BG}_1$, ${K}=22Z_c$; 
for $\mathbf{BG}_2$, ${K}=10Z_c$.
Each block therefore contains ${K}-{K}'$
filler bits.
\begin{equation}\label{eqn:cb_seg_compute}
\mathsf K_b \cdot Z_c \geq {K}^\prime
\end{equation}
\begin{table}[htbp]
\centering
\caption{Lifting-size selection parameter $\mathsf K_b$~\citep{3GPP38212}.}
\begin{tabular}{c c r}
\toprule
$\mathbf{BG}$ & $ B$     & $\mathsf K_b$ \\
\hline
1     & all   & 22 \\
2     & $B$ $>$ 640 & 10 \\
2     & 560 $<$ $B$ $\leq$ 640 & 9 \\
2     & 192 $<$ $B$ $\leq$ 560 & 8 \\
2     & $B$ $\leq$192 & 6 \\
\bottomrule
\end{tabular}%
\label{tab:kb_value}%
\end{table}%

The output of code block segmentation is calculated as follows.\\
If $C=1$,
\begin{equation}
c_{0,k}=
\begin{cases}
b_k, & 0\leq k<B,\\
\texttt{NULL}, & B\leq k<{K}.
\end{cases}
\end{equation}

If $C>1$, a CRC of length $L=24$ is attached to each
code block using the generator polynomial
\begin{equation}
\mathrm g_{\mathsf{CRC24B}}(\beta)
=\beta^{24}+\beta^{23}+\beta^6+\beta+1.
\end{equation}
Let the CRC bits for block $r$ be
$p_{r,0},p_{r,1},\ldots,p_{r,L-1}$. Then,
\begin{equation}
c_{r,k}=
\begin{cases}
b_{r({K}'-L)+k},
  & 0\leq k<{K}'-L,\\
p_{r,k+L-{K}'},
  & {K}'-L\leq k<{K}',\\
\texttt{NULL},
  & {K}'\leq k<{K},
\end{cases}
\end{equation}
where $0\leq r<C$ and ${K}$ is the encoder input
length, including filler bits.

\item {\em LDPC encoding}.\\
Each code block is encoded independently.
The input sequence is represented by
$\mathbf c_r=[c_{r,0},c_{r,1},\ldots,
c_{r,\mathrm K_r-1}]^{\mathsf T}$,
where $\mathrm K_r={K}$ is the encoder input length,
including filler bits.

The output sequence is
$d_{r,0},d_{r,1},\ldots,d_{r,\mathrm N_r-1}$,
where $\mathrm N_r=66Z_c$ for $\mathbf{BG}_1$
and $\mathrm N_r=50Z_c$ for $\mathbf{BG}_2$.
The encoding procedure is as follows~\citep{3GPP38212}.

\begin{enumerate}
\item Find the lifting-size set with index iLS
that contains $Z_c$.

\item For $k=2Z_c,\ldots,\mathrm K_r-1$,
set $d_{r,k-2Z_c}=c_{r,k}$ if
$c_{r,k}\neq\texttt{NULL}$.
Otherwise, set $c_{r,k}=0$ for parity calculation
and $d_{r,k-2Z_c}=\texttt{NULL}$ for rate matching.
All filler positions in $\mathbf c_r$, including any in the first $2Z_c$ positions, are treated as zero during parity calculation.

\item Generate $\mathrm N_r+2Z_c-\mathrm K_r$
parity bits, collected in the column vector $\mathbf w_r$, such that
\begin{equation}
\mathbf H
\begin{bmatrix}
\mathbf c_r\\
\mathbf w_r
\end{bmatrix}
=\mathbf 0.
\end{equation}
All operations are performed in $\mathbb F_2$.

\item Set
$d_{r,k-2Z_c}=w_{r,k-\mathrm K_r}$
for $k=\mathrm K_r,\ldots,\mathrm N_r+2Z_c-1$.
\end{enumerate}

\item {\em Rate matching}.\\
Rate matching adapts the number of coded bits to the
available transmission resources. It is performed
independently for each code block and consists of bit
selection and bit interleaving~\citep{3GPP38212}.

The redundancy version (RV), with index
$\mathrm{rv}_{\mathrm{id}}\in\{0,1,2,3\}$, determines
the starting position $k_0$ in a circular buffer.
This position depends on the RV index, the selected
base graph, and the circular-buffer length, as specified
in Table~5.4.2.1-2 of~\citep{3GPP38212}.
The first $2Z_c$ systematic positions have already
been omitted from the LDPC encoder output.

For the $r$th code block, the encoder output is
$d_{r,0},d_{r,1},\ldots,d_{r,\mathrm N_r-1}$.
Let $N_{\mathrm{cb}}$ denote the circular-buffer length,
where $N_{\mathrm{cb}}=\mathrm N_r$ when limited-buffer
rate matching is disabled, and
$N_{\mathrm{cb}}=\min(\mathrm N_r,N_{\mathrm{ref}})$
otherwise. The parameter $N_{\mathrm{ref}}$ is defined
in Section~5.4.2.1 of~\citep{3GPP38212}.

Let $E_r$ be the required number of output bits for
this code block. The selected sequence
$e_{r,0},e_{r,1},\ldots,e_{r,E_r-1}$ is obtained
as follows:
\begin{enumerate}
\item Initialize $k=0$ and $j=0$.
\item While $k<E_r$, inspect the buffer position
$q=(k_0+j)\bmod N_{\mathrm{cb}}$.
If $d_{r,q}\neq\texttt{NULL}$, set
$e_{r,k}=d_{r,q}$ and increment $k$.
Increment $j$ after every inspected position.
\end{enumerate}
Thus, filler positions are skipped, and the circular
buffer may be traversed repeatedly until $E_r$ bits
have been selected.

The selected bits are then interleaved according to
the modulation order $Q_m$:
\begin{equation}
f_{r,i+jQ_m}=e_{r,iE_r/Q_m+j},
\end{equation}
where $0\leq i<Q_m$ and $0\leq j<E_r/Q_m$, with $E_r$ divisible by $Q_m$.
The resulting rate-matched sequence is
$f_{r,0},f_{r,1},\ldots,f_{r,E_r-1}$.

\item {\em Code block concatenation}.\\
The rate-matched code blocks are concatenated
sequentially to form
$g_0,g_1,\ldots,g_{G-1}$, where
$G=\sum_{r=0}^{C-1}E_r$.
Let $S_r=\sum_{q=0}^{r-1}E_q$, with $S_0=0$.
Then,
\begin{equation}
g_{S_r+k}=f_{r,k},
\qquad 0\leq r<C,\quad 0\leq k<E_r.
\end{equation}
\end{enumerate}

At the receiver, the corresponding processing steps
are code block de-concatenation, rate recovery,
LDPC decoding, code block de-segmentation, and CRC
verification.
\begin{enumerate}
\item {\em Code block de-concatenation}.\\
Code block de-concatenation divides the received LLR sequence into $C$ blocks of lengths $E_r$. Let $y_0,y_1,\ldots,y_{G-1}$ denote the channel LLRs after demodulation and descrambling, ordered according to the transmitted coded-bit sequence. The LLRs associated with the $r$th rate-matched block are
\begin{equation}
\lambda^{(f)}_{r,k}=y_{S_r+k},\qquad 0\leq k<E_r,
\end{equation}
where $S_r=\sum_{q=0}^{r-1}E_q$, $S_0=0$, and $0\leq r<C$.

\item {\em Rate de-matching}.\\
Rate de-matching recovers the LLR sequence associated
with the full LDPC codeword. It is performed
independently for each code block and consists of
bit de-interleaving, circular-buffer recovery, and
restoration of the first $2Z_c$ punctured positions.

First, bit de-interleaving produces
\begin{equation}
\lambda^{(e)}_{r,iE_r/Q_m+j}
=
\lambda^{(f)}_{r,i+jQ_m},
\end{equation}
where $0\leq i<Q_m$ and $0\leq j<E_r/Q_m$, with $E_r$ divisible by $Q_m$.

Let $\pi_r(k)$ denote the circular-buffer position
from which the transmitted bit $e_{r,k}$ was selected.
This mapping is determined by the redundancy version,
the circular-buffer length, and the filler positions.
The recovered LLRs are
\begin{equation}
\Lambda_{r,v}=
\begin{cases}
+\infty,
  & v\in\mathcal F_r,\\
\displaystyle
\sum_{\substack{0\leq k<E_r\\
                2Z_c+\pi_r(k)=v}}
\lambda^{(e)}_{r,k},
  & v\notin\mathcal F_r,
\end{cases}
\end{equation}
for $0\leq v<\mathrm N_r+2Z_c$, where
$\mathcal F_r=\{{K}',\ldots,{K}-1\}$
is the set of filler positions in the full codeword.

An empty sum is zero, corresponding to an unobserved
position. Repeated observations of the same coded bit
are combined by adding their LLRs.
With the convention
$\mathscr L=\ln[\Pr(c=0)/\Pr(c=1)]$, known zero filler bits
have LLR $+\infty$, replaced by a sufficiently large
positive value in practical implementations.

\item {\em LDPC decoding}.\\
LDPC decoding is performed on each code block
independently, and different decoding algorithms
can be used. Subsection~\ref{subsection:ldpc_decoding}
describes different LDPC decoding algorithms.

\item {\em Code block de-segmentation}.\\
Code block de-segmentation is used to recover the
message bits and the transport-block CRC bits.
Assume that the output from code block
de-segmentation is
$\hat{b}(0),\hat{b}(1),\ldots,\hat{b}(B-1)$,
where $B$ is the length of the original transport
block including its CRC bits.

The input consists of the decoded encoder-input
sequences
$h_r(0),h_r(1),\ldots,h_r(\mathrm K_r-1)$,
where $\mathrm K_r={K}$ includes filler
positions and $0\leq r<C$.
When code-block CRC bits are present, they are checked before removal.
After removing the filler bits and, when present,
the code-block CRC bits, the recovered sequence
is obtained as
\begin{equation}
\hat{b}_{r({K}'-L)+s}=h_{r,s},
\end{equation}
where
$0\leq r<C$ and $0\leq s<{K}'-L$.
Here, ${K}'=B'/C$ is the number of information
and code-block CRC bits before filler insertion,
and $L$ is the code-block CRC length:
$L=0$ for $C=1$ and $L=24$ for $C>1$.
Thus, $C({K}'-L)=B$.

\item {\em CRC check}.\\
The transport-block CRC bits are extracted from
the recovered sequence. The receiver recalculates
the CRC from the decoded transport-block payload
and compares it with the decoded CRC bits.
A mismatch indicates a detected decoding error,
whereas a successful CRC check does not exclude
all possible residual errors.

\end{enumerate}

\subsection{Soft-Decision LDPC Decoding Algorithms}\label{subsection:ldpc_decoding}
Soft-decision iterative decoding involves the computation of {\em log-likelihood ratios} (LLRs) and the exchange of extrinsic information between variable nodes and check nodes. The {\em sum-product algorithm} (SPA), also known as belief propagation (BP), is a message-passing algorithm. The {\em min-sum algorithm} (MSA) approximates the SPA check-node update with lower computational complexity. Different message-update rules and decoding schedules provide trade-offs between error-correction performance and implementation complexity.

\begin{figure}
    \centering
  \includegraphics[width=0.8\linewidth]{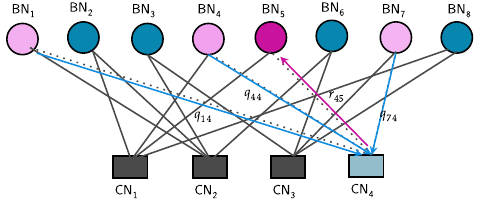}
  \caption{Check Node Updates: \scriptsize {$\mathsf{CN}_4 \longrightarrow \mathsf{BN}_5$ .}}
\label{fig:cknode_update}
\end{figure}

\begin{figure}
\centering
\includegraphics[width=0.8\linewidth]{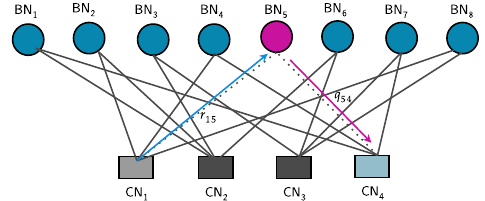}
  \caption{Bit Node Updates: \scriptsize {$\mathsf{BN}_5 \longrightarrow \mathsf{CN}_4$.}}
\label{fig:bitnode_update}
\end{figure}

Message-passing decoding consists of bit or variable node processing (column operations) and check node processing (row operations). The messages $\mathrm q_{ij}$ and $\mathrm r_{ji}$ represent variable-to-check and check-to-variable information, respectively, as illustrated in Figures~\ref{fig:cknode_update} and~\ref{fig:bitnode_update}.

\begin{figure}
\centering
\includegraphics[width=0.8\linewidth]{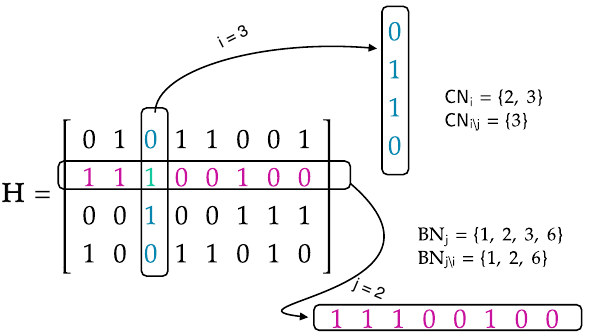}
\caption{Conceptual Illustration of CNs and BNs processing within BN and CN index sets for the specified values of $i$ and $j$.}
  \label{fig:node_index_proc}
\end{figure}

Index sets $\mathsf{BN}_{j\backslash i}$ and $\mathsf{CN}_{i\backslash j}$ are defined by the {\em parity-check matrix} (PCM). The sets $\mathsf{CN}_i=\{j:H_{j,i}=1\}$ and $\mathsf{BN}_j=\{i:H_{j,i}=1\}$ contain the indices of the check nodes connected to variable node $i$ and the variable nodes connected to check node $j$, respectively. We write $\mathsf{BN}_{j\backslash i}=\mathsf{BN}_j\setminus\{i\}$ and $\mathsf{CN}_{i\backslash j}=\mathsf{CN}_i\setminus\{j\}$.
Figure~\ref{fig:node_index_proc} provides a simple conceptual illustration of BN and CN index sets within the PCM provided in (\ref{eqn:pcm_ldpc}) for $i=3$ and $j=2$. The node labels in these illustrations use one-based indexing; the coded-bit sequences use zero-based indexing.

Furthermore, the beliefs can be described via the following equations.
Consider an LDPC codeword $\mathbf c=[{c}(0),{c}(1),\ldots,{c}({N}-1)]$. In this subsection, ${N}$ denotes the full codeword length, including punctured positions, and $\mathbf H$ has ${N}$ columns. For the NR decoder described above, this length is $\mathrm N_r+2Z_c$.

We use the following notation:
\begin{itemize}
  \item  $\mathsf{BN}_j$ = {Bit nodes connected to check node $j$},
  \item  $\mathsf{BN}_{j\backslash i}$ ={bit nodes connected to check node $j$}, excluding {bit node $i$},
  \item  $\mathsf{CN}_{i}$ = {check nodes connected to variable node $i$},
  \item  $\mathsf{CN}_{i\backslash j}$ ={check nodes connected to bit node $i$}, excluding check node $j$,
  \item  $p_i = \Pr(\mathrm c_i = 1\mid\mathrm y_i)$ is the channel-based probability that bit $\mathrm c_i$ equals 1.
  \item $\mathrm y_i$ is the channel sample at variable node $i$.
  \item The check-to-variable extrinsic message passing to variable node $i$ from check node $j$ is denoted by $\mathscr L\left(\mathrm r_{j i}\right)$ and $\mathscr L\left(\mathrm q_{i j}\right)$ is the variable-to-check extrinsic message.
\end{itemize}

Belief propagation can be implemented in the probability domain or the log domain, where binary messages are represented by LLRs. Employing LLR domain decoding offers a reduction in implementation complexity, since multiplications in the probability domain can be equivalently represented as additions in the log domain.
Besides, many multiplications of probabilities involved could become numerically unstable, so the log domain algorithm is preferred \citep{Moon2005}.

\subsubsection{Sum Product Algorithm (SPA)}~\\
The sum-product algorithm can be implemented using probabilities or LLRs~\citep{Gallager63,Moon2005}. For a transmitted LDPC codeword $\mathbf c$, the channel-based input LLR is
\begin{equation}
\mathscr L(\mathrm c_i)=\ln\frac{\Pr(\mathrm c_i=0\mid\mathrm y_i)}{\Pr(\mathrm c_i=1\mid\mathrm y_i)}.
\end{equation}
Here, the probability uses the channel observation and the assumed bit prior, before incorporating the parity-check constraints. At the rate-recovery output, unobserved punctured positions have zero LLR, and known zero filler positions are assigned a large positive LLR in practice.

The messages $\mathrm q_{ij}(b)$ and $\mathrm r_{ji}(b)$ are probability-domain messages for $b\in\{0,1\}$, and their LLR representations are $\mathscr L(\mathrm q_{ij})=\ln[\mathrm q_{ij}(0)/\mathrm q_{ij}(1)]$ and $\mathscr L(\mathrm r_{ji})=\ln[\mathrm r_{ji}(0)/\mathrm r_{ji}(1)]$.
On a Tanner graph with cycles, the resulting posterior LLRs are generally approximations. In each iteration, the algorithm updates its key components through horizontal and vertical processing steps.

The check-to-bit-node operation ({\em horizontal processing}) is based on (\ref{eqn:cknod_update}).
\begin{equation}\label{eqn:cknod_update}
\begin{aligned}
&\mathscr L\left(\mathrm r_{j i}\right)=\ln \frac{\mathrm r_{j i}(0)}{\mathrm r_{j i}(1)}\\&=2 \tanh ^{-1}\left(\prod_{i^{\prime} \in \mathsf{BN}_{j\backslash i}} \tanh \left(\frac{1}{2} \mathscr L\left(\mathrm q_{i^{\prime} j}\right)\right)\right)
\\&=\left(\prod_{i^{\prime} \in \mathsf{BN}_{j\backslash i}}\operatorname{sign}\left(\mathscr L\left(\mathrm q_{i^{\prime} j}\right)\right)\right) \phi\left(\sum_{i^{\prime} \in \mathsf{BN}_{j\backslash i}} \phi\left(\left|\mathscr L\left(\mathrm q_{i^{\prime} j}\right)\right|\right)\right).
\end{aligned}
\end{equation}

Here, $\phi(x)=-\ln[\tanh(x/2)]=\ln[(\mathrm e^x+1)/(\mathrm e^x-1)]$ for $x>0$, with $\phi(0)=+\infty$ by continuity.

For sign products, take $\operatorname{sign}(0)=+1$; a zero incoming magnitude still gives a zero outgoing magnitude. The hyperbolic-tangent and $\phi$ expressions are interpreted by limits at zero or infinity. Practical implementations use finite LLR saturation and stable check-node computations.

The bit-to-check-node operation ({\em vertical processing}) is given by
\begin{equation}\label{eqn:bitnod_update}
\mathscr L\left(\mathrm q_{i j}\right)=\mathscr L\left(\mathrm c_i\right)+\sum_{j^{\prime} \in \mathsf{CN}_{i\backslash j} } \mathscr L\left(\mathrm r_{j^{\prime} i}\right).
\end{equation}


\begin{equation}
\mathscr L\left(Q_i\right)=\mathscr L\left(\mathrm c_i\right)+\sum_{j^{\prime} \in \mathsf{CN}_{i}} \mathscr L\left(\mathrm r_{j^{\prime} i}\right).
\end{equation}

The estimated posterior LLR $\mathscr L(Q_i)$ is used to make a hard decision:

\begin{equation}
\hat{{c}}_i=\left\{\begin{array}{lc}
1, & \text { if } \mathscr L\left(Q_i\right)<0 ,\\
0, & \text { else .}
\end{array}\right.
\end{equation}

Repeat these steps until the maximum number of iterations is reached or $\mathbf H\hat{\mathbf c}^{\mathsf T}=\mathbf0$. A zero syndrome indicates that the estimate satisfies the parity checks; it does not guarantee that the transmitted codeword has been recovered. For the flooding schedule below, initialize $\mathscr L(\mathrm q_{ij})=\mathscr L(\mathrm c_i)$ on every edge. The check-node messages are updated first, followed by the variable-node messages.
%
%
\definecolor{lightgray}{gray}{0.5}
\begin{algorithm}[hbt!]
\caption{\small{Log-domain BP decoding (flooding schedule)}}
\label{alg:communication}\label{fig:algo_ldpc_dec_spa}
\KwIn{Channel LLRs $\mathscr L_i$, parity-check matrix $\mathbf H$, maximum iteration count $I_{\max}\geq1$.}
\KwOut{Estimated codeword $\hat{\mathbf c}$ and parity-check status.}
Initialize $\mathscr L(q_{ij})\gets\mathscr L_i$ for every edge $(j,i)$\;
\For{$t\gets1$ \KwTo $I_{\max}$}{
\For{each check node $j$}{
\For{each $i\in\mathsf{BN}_j$}{
Compute $\mathscr L(r_{ji})$ using (\ref{eqn:cknod_update})\;
}}
\For{each variable node $i$}{
$\mathscr L(Q_i)\gets\mathscr L_i+\sum_{j\in\mathsf{CN}_i}\mathscr L(r_{ji})$\;
$\hat c_i\gets\mathbf1\{\mathscr L(Q_i)<0\}$\;
\For{each $j\in\mathsf{CN}_i$}{
$\mathscr L(q_{ij})\gets\mathscr L_i+\sum_{j^{\prime}\in\mathsf{CN}_{i\backslash j}}\mathscr L(r_{j^{\prime}i})$\;
}}
\If{$\mathbf H\hat{\mathbf c}^{\mathsf T}=\mathbf0$}{
\Return{$\hat{\mathbf c}$, parity checks satisfied}\;
}}
\Return{$\hat{\mathbf c}$, maximum iterations reached}\;
\end{algorithm}
 The BP algorithm can achieve strong decoding performance, but exact check-node updates have substantial computational cost. In order to find a better trade-off between performance and complexity, a number of efficient decoding algorithms have been proposed in the scientific literature.

\subsubsection{Min-Sum Algorithm (MSA)}~\\
The min-sum algorithm (MSA) reduces the complexity of the SPA check-node update. Since $\phi(x)$ decreases toward zero as $x$ increases, the largest term in the sum below corresponds to the smallest incoming LLR magnitude. MSA approximates the sum by this largest term:
\begin{equation}
\sum_{i^{\prime} \in \mathsf{BN}_{j\backslash i}} \phi\left(\left|\mathscr L\left(\mathrm q_{i^{\prime} j}\right)\right|\right) \approx \phi\left(\min _{i^{\prime} \in \mathsf{BN}_{j\backslash i}}\left(\left|\mathscr L\left(\mathrm q_{i^{\prime} j}\right)\right|\right)\right).
\end{equation}

Using $\phi(\phi(x))=x$ for $x>0$ yields the MSA check-node update below, with limiting values at zero. The variable-node updates and hard decisions remain the same as in SPA.

\begin{equation}\label{eqn:msa_algo}
\mathscr L\left(\mathrm r_{j i}\right)=\left(\prod_{i^{\prime} \in \mathsf{BN}_{j\backslash i}}\operatorname{sign}\left(\mathscr L\left(\mathrm q_{i^{\prime} j}\right)\right)\right) \cdot \min _{i^{\prime} \in \mathsf{BN}_{j\backslash i}}\left(\left|\mathscr L\left(\mathrm q_{i^{\prime} j}\right)\right|\right).
\end{equation}

\subsubsection{Optimized Min-Sum Decoding Algorithm}~\\
For the same incoming messages, the MSA check-to-variable output magnitude is greater than or equal to the SPA magnitude: replacing the sum of positive $\phi$ terms by its largest term decreases the argument of the decreasing function $\phi$. Normalized and offset min-sum decoding compensate for this overestimation by reducing check-to-variable message magnitudes~\citep{Jinghu2005}.
\begin{enumerate}[label=(\alph*)]
\item Normalized Min-Sum Algorithm, (\ref{eqn:msa_algo}) simply becomes:
\begin{equation}
\begin{aligned}
\mathscr L_{\texttt{\tiny NMSA}}\left(\mathrm r_{j i}\right)&=\left(\prod_{i^{\prime} \in \mathsf{BN}_{j\backslash i}}\operatorname{sign}\left(\mathscr L\left(\mathrm q_{i^{\prime} j}\right)\right)\right) \\&\cdot \min _{i^{\prime} \in \mathsf{BN}_{j\backslash i}}\left(\alpha\cdot\left|\mathscr L\left(\mathrm q_{i^{\prime} j}\right)\right|\right),
\end{aligned}
\end{equation}
where $\alpha\in(0,1)$ is a normalization or scaling factor common to the incoming magnitudes in this check-node update.

\item Offset Min-Sum Algorithm, (\ref{eqn:msa_algo})  becomes:
\begin{equation}
\begin{aligned}
\mathscr L_{\texttt{\tiny OMSA}}\left(\mathrm r_{j i}\right)=&\left(\prod_{i^{\prime} \in \mathsf{BN}_{j\backslash i}}\operatorname{sign}\left(\mathscr L\left(\mathrm q_{i^{\prime} j}\right)\right)\right) \\&\cdot \max\left(\min _{i^{\prime} \in \mathsf{BN}_{j\backslash i}}\left(\left|\mathscr L\left(\mathrm q_{i^{\prime} j}\right)\right|-\beta\right), 0 \right),
\end{aligned}
\end{equation}
 where $\beta \geq 0$ is the offset value.
\end{enumerate}

\subsubsection{Layered Belief Propagation Algorithm}~\\
The {\em layered belief propagation} (\acs{LBP}) algorithm processes subsets of rows ({\em layers}) of the PCM sequentially~\citep{Hocevar2004}. Unlike flooding BP, it makes updated LLRs available immediately to subsequent check-node updates.

Before the first iteration, initialize $\mathscr L(\mathrm Q_i)=\mathscr L(\mathrm c_i)$ for every variable node and $\mathscr L(\mathrm r_{ji})=0$ for every edge. One iteration consists of a complete pass through all layers.

For a check node $j$ and each connected variable node $i\in\mathsf{BN}_j$, the check-node processing (CNP) uses the following steps:
\begin{equation}
\begin{array}{l}
\texttt{(1) Update input LLRs:}\\
\quad\mathscr L(\mathrm q_{ij})=
\mathscr L(\mathrm Q_i)-\mathscr L(\mathrm r_{ji}).\\[2pt]
\texttt{(2) Perform CNP:}\\
\quad\mathscr L^{\prime}(\mathrm r_{ji})=
2\tanh^{-1}\!\left(
\displaystyle\prod_{i^{\prime}\in\mathsf{BN}_{j\backslash i}}
\tanh\!\left(\frac{\mathscr L(\mathrm q_{i^{\prime}j})}{2}\right)
\right).\\[2pt]
\texttt{(3) Update output LLRs:}\\
\quad\mathscr L^{\prime}(\mathrm Q_i)=
\mathscr L(\mathrm q_{ij})+
\mathscr L^{\prime}(\mathrm r_{ji}).
\end{array}
\end{equation}
Here, unprimed quantities are the values available before processing check node $j$, and primed quantities are the newly computed values. Step (1) removes the old contribution of check node $j$ from the current posterior estimate, and step (3) inserts its new contribution. All incoming messages for check node $j$ must be computed before its updated output LLRs are stored.

After this check-node update, replace
$\mathscr L(\mathrm r_{ji})\gets\mathscr L^{\prime}(\mathrm r_{ji})$
and
$\mathscr L(\mathrm Q_i)\gets\mathscr L^{\prime}(\mathrm Q_i)$
for all $i\in\mathsf{BN}_j$. The next check node therefore uses the most recent available LLRs. Posterior LLRs for variable nodes not connected to $j$ remain unchanged.

Check nodes within a layer can be processed in parallel without conflicting posterior updates when they do not share variable nodes; equivalently, the corresponding layer submatrix has at most one nonzero entry per column. Otherwise, these check nodes must be processed sequentially for the update rule above, or a schedule with explicit conflict handling must be used. Successive layers can share variable nodes and are not independent.

The benefit of layered scheduling comes from the immediate reuse of updated information. It can reduce the number of iterations required to achieve a given error rate compared with flooding BP~\citep{Hocevar2004,Zhang2014}. The reduction depends on the code, schedule, and operating conditions, and does not imply a universal factor-of-two reduction in decoding time.

After a complete pass through all layers, use the current $\mathscr L(\mathrm Q_i)$ values to make hard decisions and evaluate $\mathbf H\hat{\mathbf c}^{\mathsf T}$. Stop when all parity checks are satisfied or the maximum number of iterations is reached. The final LLRs are the stored posterior estimates for all variable nodes, including those not updated by the last layer.


\section{5G NR Polar Codes}
Polar codes were introduced by \citet{Arikan2009} in a paper published in 2009. They quickly attracted attention because of their explicit construction, capacity-achieving asymptotic behavior, and low-complexity encoding and decoding. Their adoption for 5G NR control and broadcast information demonstrates their practical relevance~\citep{3GPP38212}.

Polar codes achieve the symmetric capacity of binary-input discrete memoryless channels asymptotically under {\em successive cancellation} (\acs{SC}) decoding, with encoding and decoding complexity $\mathcal O(N\log N)$~\citep{Arikan2009}. This result concerns sequences of codes as the block length increases and does not imply capacity-achieving performance at a fixed finite length. For symmetric channels, the symmetric capacity equals the channel capacity.

Polar coding is based on channel polarization, code construction, and a binary linear transform. Information is assigned to reliable synthesized channels, while the remaining input positions are frozen to known values. With zero frozen bits, a polar code is a linear block code; fixed nonzero frozen bits generally define an affine coset. The standard binary construction uses a mother-code length $N=2^n$, while rate matching provides flexibility in the transmitted length and effective rate. In 5G NR, these properties are used for short control and broadcast messages, for which finite-length error performance and decoding latency are important.

\subsection{State-of-the-Art Polar Codes}
Polar codes have emerged as a key channel coding scheme for 5G NR \citep{Bioglio2021}. The trend in polar code design and decoding techniques has been driven by the need for efficient and reliable communication. These codes exhibit promising characteristics such as rate flexibility and low decoding latency, addressing crucial requirements for 5G systems. However, challenges persist, particularly in reducing latency without compromising reliability \citep{Gamage2020}. Efforts have been made to enhance the decoding process, such as the development of low latency decoders for short block length polar codes. Thus, \citet{Gamage2020} highlighted decoders that utilize simplified \acs{SC} algorithms combined with list decoding techniques to achieve the desired balance between reliability and latency in ultra-reliable low-latency communication systems. Subsequently, \citet{Geiselhart2020} leveraged CRC codes to aid belief propagation list decoding, improving error-rate performance while optimizing decoding complexity. Meanwhile, \citet{Piao2019} proposed innovative decoding algorithms like CRC-aided sphere decoding to enhance the performance of short polar codes. By utilizing CRC information, these algorithms provide stable performance across various code rates. Additionally, \citet{Cavatassi2019} introduced asymmetric coding schemes to allow for arbitrary block lengths, reducing decoding complexity while maintaining error correction performance. \citet{Shen2022} explored fast iterative soft-output list decoding to improve error-rate performance and decoding efficiency. \citet{Kaykac2020} highlighted that understanding the operation and performance of 5G polar codes is crucial as they are integral to the functionality of 5G control channels. \citet{Ercan2020} optimized practical implementations of polar code decoders, developing dynamic SC-flip decoding algorithms with reduced complexity. \citet{Kestel2020} quantified trade-offs between error-correction capability and implementation costs, crucial for achieving efficient high-throughput decoding in 5G systems. Moreover, \citet{ArliGazi2021} surveyed belief-propagation decoding methods for polar codes, while \citet{Sun2019} proposed a low-complexity decoding scheme for the 5G eMBB control channel.

Moreover, the advancements in polar code decoding algorithms and software and hardware architectures have paved the way for efficient and low-latency implementations, including 5G and beyond. \citet{Sarkis2014a} introduced a framework for generating high-speed software polar decoders, achieving significant throughput improvements. Meanwhile, in  \citep{Sarkis2014b}, simplified decoding algorithms have been proposed to enhance the speed of polar list decoders, maintaining error-correction performance. Subsequently, \citet{Kam2021} addressed the latency issue in SC decoding by introducing tree-level parallelism and novel pruning methods, significantly reducing decoding latency. Moreover, \citet{Xiang2020} presented a reduced-complexity logarithmic SC stack (Log-SCS) polar decoding algorithm, achieving notable improvements in decoding latency and complexity. Additionally, \citet{Rezaei2022} focused on implementing ultra-fast polar decoders with new sub-codes and decoding algorithms for short to moderate block lengths, emphasizing hardware optimization techniques. Furthermore, \citet{Liu2022} proposed high-throughput adaptive list decoding architectures for polar codes on GPUs, leveraging adaptive mapping strategies to improve throughput and latency performance.

Recent research combines deep learning, reinforcement learning, and traditional decoding methods to improve the efficiency and latency of polar decoding. \citet{Doan2019a} introduced a novel approach, {\em neural belief propagation} (NBP), combining CRC with polar codes to enhance error-correction performance, particularly for parallel iterative BP decoders. Building upon this, \citet{Hashemi2019} proposed a deep-learning-aided successive-cancellation list (DL-SCL) decoding algorithm, leveraging deep learning techniques to optimize bit-flipping metrics and reduce computational complexity. Meanwhile, \citet{Doan2020} addressed factor-graph permutation selection in polar codes using reinforcement learning, achieving significant error-correction performance gains and expanding on this framework, fast SC-flip decoding, a bit-flipping algorithm optimized via reinforcement learning, was introduced by \citet{Doan2021} for improved error-correction performance in polar codes. Additionally, \citet{Wodiany2019} presented a low-precision {\em neural network} (NN) decoder to mitigate the scalability issues and high memory usage of conventional NN decoders, reducing weight storage and memory traffic at the cost of a small error-performance loss. In parallel, \citet{Wen2019} proposed a BP-NN decoding algorithm for polar codes, integrating neural network decoders into the belief propagation framework to reduce decoding delay while maintaining low bit error rates. These advancements in decoding algorithms, combining techniques from deep learning, reinforcement learning, and traditional decoding methods, offer promising solutions for efficient and low-latency decoding of polar codes in future communication systems.

Furthermore, the primary polar code decoding algorithms include SC~\citep{Arikan2009}, SCL~\citep{Stimming2014,Tal_vardy2015}, CRC-aided SCL (CA-SCL)~\citep{Zhang2017}, BP~\citep{Arikan2008}, and soft cancellation (SCAN)~\citep{Fayyaz2014}. SC decoding has limited finite-length performance, while SCL retains multiple candidate paths and CA-SCL uses CRC checks to help select among them. CA-SCL is widely studied for NR polar codes, but TS 38.212 specifies the encoding procedures and does not mandate this decoding algorithm~\citep{3GPP38212}.

Conventional SC, SCL, and CA-SCL implementations provide hard decisions. BP and SCAN can provide soft outputs for iterative receiver processing~\citep{Fayyaz2013,Fayyaz2014}. BP commonly uses flooding message passing, whereas SCAN follows a sequential schedule related to SC~\citep{Yuan2014,Fayyaz2014}. 

Their latency, convergence, and error-correction performance
depend on the code, stopping rule, iteration count, and
implementation. Subject to these qualifications, the following
qualitative ranking may be considered for the configurations
and error-rate criteria:
 $\texttt{CA-SCL} > \texttt{state of the art (LDPC, Turbo codes)} > \texttt{SCL} > \texttt{BP} = \texttt{SCAN} > \texttt{SC}$.
Here, $>$ denotes better error-rate performance under the
specified comparison conditions. This illustrative ordering
is not universal and does not imply equivalent decoding
complexity or latency.
\subsection{Foundations and Fundamentals}
A polar code of length $N=2^n$ is constructed from an $N\times N$ polar transform matrix $\mathbf G_N$. Let $\mathcal A\subseteq\{0,\ldots,N-1\}$ be the information set, with $|\mathcal A|=K$. The input vector $\mathbf u$ contains $K$ information bits in $\mathcal A$ and $N-K$ zero frozen bits in $\mathcal A^c$. The codeword is $\mathbf x=\mathbf u\mathbf G_N$, with all operations performed in $\mathbb F_2$. The $K$ rows of $\mathbf G_N$ indexed by $\mathcal A$ form a $K\times N$ generator matrix for this code.

Using Arikan's bit-reversed convention, the transform matrix is
\begin{equation}\label{eqn:gen_polar}
\mathbf {G}_{N}=\mathbf B_{N} \mathbf F^{\otimes n},
\end{equation}
where $\mathbf B_{N}$ is the bit-reversal permutation matrix, $\mathbf F^{\otimes n}$ is the $n$-fold Kronecker product of $\mathbf F$ with itself, defined recursively for $n\geq2$ as
\begin{equation}
\begin{array}{@{}c@{\;}c@{}}
\mathbf F^{\otimes 1} = \mathbf F=\left[\begin{array}{ll}
1 & 0 \\
1 & 1
\end{array}\right], & \mathbf F^{\otimes n}=\left[\begin{array}{cc}
\mathbf F^{\otimes(n-1)} & \mathbf {0} \\
\mathbf F^{\otimes(n-1)} & \mathbf F^{\otimes(n-1)}
\end{array}\right].
\end{array}
\end{equation}
The encoding operation is therefore
\begin{equation}
x_0^{N-1}=\left(u_0^{N-1}\right) \mathbf B_{N} \mathbf F^{\otimes n}.
\end{equation}

\begin{figure}[ht]
\centering
\includegraphics[width=0.8\linewidth]{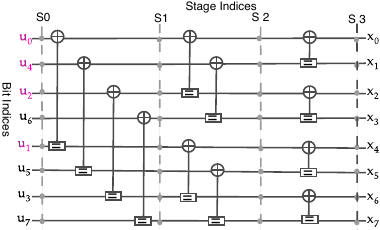}
\caption{An $(8,4)$ polar encoder with bit-reversed indexing.}
\label{fig:enc_polar}
\end{figure}

The {\em bit-reversal permutation matrix} $\mathbf B_N$ does not change the code's distance properties when the information and frozen sets are permuted consistently. Some implementations use the {\em natural-order} transform $\widetilde{\mathbf G}_N=\mathbf F^{\otimes n}$. The indexing convention must be used consistently in encoding, code construction, and decoding~\citep{Arikan2009,Bioglio2021}.
For instance, consider a polar code of length $N=8$. Once the four zero frozen positions (magenta) and the four information positions (dark) are determined, the construction defines an $(8,4)$ code.
The encoder graph is depicted in Figure~\ref{fig:enc_polar}. An interactive illustration of polar encoding is available in~\citep{Elkelesh}.
Consequently, the input/output relationship  in (\ref{eqn:io_polar_graph_enc}) is established according to the  encoder graph representation shown in Figure \ref{fig:enc_polar}.
\begin{equation}\label{eqn:io_polar_graph_enc}
\begin{array}{ll}
\begin{aligned}
x_0&=\bigoplus_{j=0}^{7}u_j,\\
x_4&=u_1\oplus u_3\oplus u_5\oplus u_7,\\
x_2&=u_2\oplus u_3\oplus u_6\oplus u_7,\\
x_6&=u_3\oplus u_7,
\end{aligned}
&\begin{aligned}
x_1&=u_4\oplus u_5\oplus u_6\oplus u_7,\\
x_5&=u_5\oplus u_7,\\
x_3&=u_6\oplus u_7,\\
x_7&=u_7,
\end{aligned}
\end{array}
\end{equation}
where $\oplus$ denotes addition in $\mathbb F_2$.
The codeword $\left[x_0, x_1, \ldots, x_7\right]$ is transmitted through eight independent uses of a binary-input memoryless channel $W$, with channel outputs $\left[y_0, y_1, \ldots, y_7\right]$. Their combined transition probability is $W^8(\mathbf y\mid\mathbf x)=\prod_{i=0}^{7}W(y_i\mid x_i)$. Combining this channel with the polar transform gives $W_8(\mathbf y\mid\mathbf u)=W^8(\mathbf y\mid\mathbf u\mathbf G_8)$, from which the synthesized bit-channels are defined.
Initially, polar codes were nonsystematic, but they can also be encoded systematically. Systematic polar encoding uses the same underlying polar code. Under the corresponding SC decoding procedure, it preserves the frame-error performance and can improve the bit-error performance compared with nonsystematic encoding~\citep{Arikan2011Systematic}. These comparisons depend on the decoding and output-recovery procedures.

In a linear code, a codeword lies in the row space of its generator matrix. Let $\mathbf G_{\mathcal A}$ be the $K\times N$ generator matrix formed by the information rows of the polar transform. In $\mathbf x=\mathbf z\mathbf G_{\mathcal A}$, $\mathbf z\in\mathbb F_2^K$ is an encoding-input vector and $\mathbf x$ is the codeword. To encode a message $\boldsymbol m$ systematically, choose $K$ codeword positions $\mathcal S$ such that the square submatrix $\mathbf G_{\mathcal A,\mathcal S}$ is invertible, and set
\begin{equation}
\mathbf z=\boldsymbol m\mathbf G_{\mathcal A,\mathcal S}^{-1},
\qquad \mathbf x=\mathbf z\mathbf G_{\mathcal A}.
\end{equation}
Then $\mathbf x_{\mathcal S}=\boldsymbol m$, so the message symbols appear explicitly in the codeword. Efficient systematic polar encoding methods are described in~\citep{Arikan2011Systematic}.

For reference, the full transform in natural order is
\begin{equation}
\widetilde{\mathbf G}_8=\mathbf F^{\otimes 3}=\left[\begin{array}{llllllll}
1 & 0 & 0 & 0 & 0 & 0 & 0 & 0 \\
1 & 1 & 0 & 0 & 0 & 0 & 0 & 0 \\
1 & 0 & 1 & 0 & 0 & 0 & 0 & 0 \\
1 & 1 & 1 & 1 & 0 & 0 & 0 & 0 \\
1 & 0 & 0 & 0 & 1 & 0 & 0 & 0 \\
1 & 1 & 0 & 0 & 1 & 1 & 0 & 0 \\
1 & 0 & 1 & 0 & 1 & 0 & 1 & 0 \\
1 & 1 & 1 & 1 & 1 & 1 & 1 & 1
\end{array}\right] .
\end{equation}

\subsection{3GPP 5G Polar Codes}
In NR, polar coding is used for the {\em broadcast channel} (BCH),
{\em downlink control information} (DCI), and {\em uplink control information} (UCI). For UCI, polar coding applies when the payload size satisfies $A\geq12$; smaller payloads use the small-block coding procedures specified
in TS 38.212~\citep{3GPP38212}.

\begin{figure*}[!ht]
\centering
\subfloat[\scriptsize{Transmitter end.}\label{fig:tx_polar}]{
\includegraphics[width=0.8\linewidth]{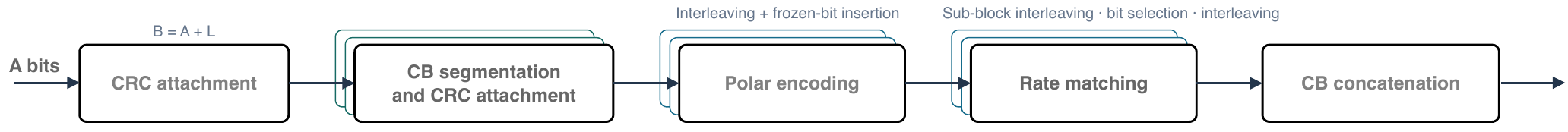}
}
\\
\subfloat[\scriptsize{Receiver end.}\label{fig:rx_polar}]{
\includegraphics[width=0.8\linewidth]{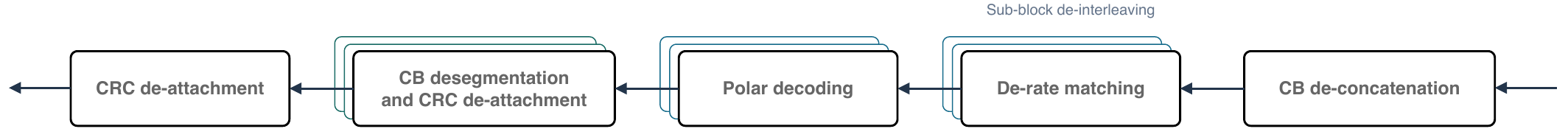}
}
\caption{Conceptual illustration of the NR polar transceiver chain. For UCI, the first stage passes the payload unchanged and CRC attachment follows segmentation. For BCH/DCI, CRC attachment is part of payload preprocessing and the UCI segmentation stage is bypassed.}
\label{fig:polar_tranceiver_chain}
\end{figure*}

Figure~\ref{fig:polar_tranceiver_chain} provides a conceptual view of polar transmission and reception. BCH, DCI, and UCI require different preprocessing. In particular, UCI segmentation, when enabled, precedes per-block CRC attachment. BCH and DCI use their own payload processing and CRC procedures; the stages below do not represent an identical preprocessing chain for all three cases.

For downlink polar coding, input interleaving distributes CRC bits among payload bits and can support partial CRC checks during decoding. These CRC bits must be distinguished from the additional polar parity-check (PC) bits used for certain short UCI payloads. A failed partial check can eliminate a candidate path; early termination of list decoding requires that no candidate path remains valid. The CRC polynomials and interleavers are specified by the standard and are not free design choices in a compliant NR implementation.

\begin{enumerate}
\item {\em CRC attachment}.\\
Let $a(0),a(1),\ldots,a(A-1)$ denote a payload of length $A$, and let $p(0),p(1),\ldots,p(L-1)$ denote its CRC bits. For BCH and DCI, $L=24$ and the generator polynomial is
\begin{equation}
\begin{array}{r}
\mathrm g_{\mathsf{CRC24C}}(\beta)=\left[\beta^{24}+\beta^{23}+\beta^{21}+\beta^{20}\right.\\
\left.+\beta^{17}+\beta^{15}+\beta^{13}+\beta^{12}+\beta^8\right.\\
\left.+\beta^4+\beta^2+\beta+1\right].
\end{array}
\end{equation}
For BCH, $a$ denotes the payload after the channel-specific preprocessing in Section~7.1 of~\citep{3GPP38212}. For DCI, CRC calculation includes the prescribed prefix of 24 ones, and the final 16 CRC bits are subsequently masked with the RNTI as specified in Section~7.3.2.

Before DCI CRC masking, the CRC-appended sequence is
\begin{equation}
b_k=\left\{\begin{array}{ll}
a_k,& 0\leq k<A,\\
p_{k-A},& A\leq k<A+L,
\end{array}\right.
\end{equation}
with $B=A+L$. For BCH, $c_{0,k}=b_k$. For DCI,
\begin{equation}
c_{0,k}=\left\{\begin{array}{@{}l@{\;}l@{}}
b_k,& 0\leq k<A+8,\\
b_k\oplus x_{\mathrm{RNTI},k-A-8},& A+8\leq k<A+24,
\end{array}\right.
\end{equation}
where $x_{\mathrm{RNTI},0},\ldots,x_{\mathrm{RNTI},15}$ are the RNTI bits in the order specified by the standard. Both cases have $C=1$ and $K_0=A+24$.

For polar-coded UCI, CRC attachment is performed per segment as described in the next step. The CRC length is $L=6$ for $12\leq A\leq19$ and $L=11$ for $A\geq20$, using
\begin{equation}
\begin{array}{ll}
\mathrm g_{\mathsf{CRC11}}(\beta)
=\left[\beta^{11}+\beta^{10}+\beta^9+\beta^5+1\right],\\
\mathrm g_{\mathsf{CRC6}}(\beta)
=\left[\beta^6+\beta^5+1\right].
\end{array}
\end{equation}

\item {\em Code block segmentation and code block CRC attachment}.\\
This step applies to polar-coded UCI. The input is $a(0),a(1),\ldots,a(A-1)$, with $12\leq A\leq1706$. Let $E$ denote the UCI rate-matching length used in the segmentation decision. The number of code blocks is
\begin{equation}
C=\left\{\begin{array}{ll}
2,& \texttt{\footnotesize{if}} (A\geq360\text{ and }E\geq1088)\\
 & \quad\text{or }A\geq1013,\\
1,& \texttt{\footnotesize{otherwise}}.
\end{array}\right.
\end{equation}
The total payload length after zero padding is
\begin{equation}
A'=\left\lceil A/C\right\rceil\cdot C.
\end{equation}
Thus, $A'$ is a padded total length, not the maximum code block size. Define
\begin{equation}
a'_k=\left\{\begin{array}{ll}
0,& 0\leq k<A'-A,\\
a_{k-(A'-A)},& A'-A\leq k<A'.
\end{array}\right.
\end{equation}
For $0\leq r<C$, the CRC bits $p_{r,0},\ldots,p_{r,L-1}$ are calculated from the $A'/C$ payload bits assigned to segment $r$. The encoder input is
\begin{equation}
c_{r,k}=\left\{\begin{array}{ll}
a'_{rA'/C+k},& 0\leq k<A'/C,\\
p_{r,k-A'/C},& A'/C\leq k<K_r,
\end{array}\right.
\end{equation}
where $K_r=A'/C+L$. Unlike LDPC segmentation, each polar-coded UCI segment receives a CRC even when $C=1$.

\item {\em Polar code length selection}.\\
Each block $c_r(0),c_r(1),\ldots,c_r(K_r-1)$
is encoded independently.
Let $E_r$ be its rate-matching output length
and $N_r=2^n$ its mother-code length.

For polar-coded UCI, the per-block rate-matching
length is $E_r=\lfloor E/C\rfloor$, according to
the applicable UCI rate-matching procedure.
For BCH/DCI, $C=1$ and $E_0=E$.

The exponent is determined by
\texttt{\footnotesize{if}} $E_r\leq(9/8)\cdot
2^{\left(\left\lceil\log_2E_r\right\rceil-1\right)}$
\texttt{\footnotesize{and}} $K_r/E_r<9/16$,\\
$\quad n_1=\left\lceil\log_2E_r\right\rceil-1$.\\
\texttt{\footnotesize{else}}
$n_1=\left\lceil\log_2E_r\right\rceil$.\\
$R_{\min}=1/8$;\quad
$n_2=\left\lceil\log_2\left(K_r/R_{\min}\right)\right\rceil$;\\
$n=\max\left\{\min\left\{n_1,n_2,n_{\max}\right\},n_{\min}\right\}$.\\
Here, $n_{\min}=5$, while $n_{\max}=9$ for BCH/DCI and $n_{\max}=10$ for polar-coded UCI. These parameters bound the exponent and hence the code length. The configuration must satisfy $K_r+n_{\mathrm{PC}}\leq E_r$, where $n_{\mathrm{PC}}$ counts additional polar PC bits and is not included in $K_r$.

\begin{itemize}
\item {\em Input interleaving}.\\
The input sequence is interleaved as
\begin{equation}
c'_{r,k}=c_{r,\Pi(k)},\qquad 0\leq k<K_r,
\end{equation}
where $\Pi(k)$ is specified in Section~5.3.1.1 of~\citep{3GPP38212}. Input interleaving is enabled for BCH/DCI ($I_{\mathrm{IL}}=1$) and disabled for UCI ($I_{\mathrm{IL}}=0$); in the latter case, $\Pi(k)=k$.

\item {\em Polar encoding}.\\
The nonfrozen set contains $K_r+n_{\mathrm{PC}}$ positions: $K_r$ positions carry the interleaved input bits and $n_{\mathrm{PC}}$ positions carry additional parity checks. The remaining positions of $\mathbf u_r=[u_{r,0},\ldots,u_{r,N_r-1}]$ are frozen to zero. The allocation uses the standardized reliability sequence and the exclusions imposed by rate matching, according to Sections~5.3.1.2 and~5.4.1.1 of~\citep{3GPP38212}.

For BCH/DCI, $n_{\mathrm{PC}}=0$. For polar-coded UCI, $n_{\mathrm{PC}}=3$ when $18\leq K_r\leq25$ and $n_{\mathrm{PC}}=0$ when $K_r>30$. The PC-position selection and five-stage shift-register calculation, including the parameter $n_{\mathrm{PC}}^{\mathrm{wm}}$, follow the standard.

NR uses the natural-order transform, without a separate bit-reversal matrix:
\begin{equation}
\mathbf d_r=\mathbf u_r\widetilde{\mathbf G}_{N_r},
\qquad\widetilde{\mathbf G}_{N_r}=\mathbf F^{\otimes n}.
\end{equation}
All operations are performed in $\mathbb F_2$. The output is $d_r(0),d_r(1),\ldots,d_r(N_r-1)$.
\end{itemize}

\item {\em Rate matching}.\\
Polar rate matching is performed per code block and consists of sub-block interleaving, bit selection, and, when enabled, coded-bit interleaving. The input is $d_r(0),\ldots,d_r(N_r-1)$, and the output is $f_r(0),\ldots,f_r(E_r-1)$.
\begin{itemize}
\item {\em Sub-block interleaving}.\\
The coded bits are divided into 32 sub-blocks of length $N_r/32$. With the permutation $P$ in Table~5.4.1.1-1 of~\citep{3GPP38212},
\begin{equation}
\begin{aligned}
J(k)&=P\!\left(\left\lfloor\frac{32k}{N_r}\right\rfloor\right)
\frac{N_r}{32}+(k\bmod(N_r/32)),\\
s_{r,k}&=d_{r,J(k)},\qquad 0\leq k<N_r.
\end{aligned}
\end{equation}
\item {\em Bit selection}.\\
For $0\leq k<E_r$, the selected sequence is
\begin{equation}
e_{r,k}=\left\{\begin{array}{@{}l@{\;}l@{}}
s_{r,k\bmod N_r},&  E_r\geq N_r,\\
s_{r,k+N_r-E_r},&  E_r<N_r,\ K_r/E_r\leq7/16,\\
s_{r,k},&  E_r<N_r,\ K_r/E_r>7/16.
\end{array}\right.
\end{equation}
These cases correspond to repetition (or no length change when $E_r=N_r$), puncturing, and shortening, respectively. Shortening also requires the compatible frozen-set construction prescribed by the standard.
\item {\em Interleaving of coded bits}.\\
Coded-bit interleaving is disabled for BCH/DCI ($I_{\mathrm{BIL}}=0$), so $f_{r,k}=e_{r,k}$. It is enabled for UCI ($I_{\mathrm{BIL}}=1$), using the triangular interleaver in Section~5.4.1.3 of~\citep{3GPP38212}. This interleaver is distinct from the input interleaver above. The rate-matching output length satisfies $E_r\leq8192$.
\end{itemize}

\item {\em Code block concatenation}.\\
The rate-matched blocks are concatenated sequentially. Let $S_r=\sum_{q=0}^{r-1}E_q$, with $S_0=0$, and $G=\sum_{r=0}^{C-1}E_r$. The concatenated sequence satisfies
\begin{equation}
g_{S_r+k}=f_{r,k},\qquad0\leq r<C,\quad0\leq k<E_r.
\end{equation}
For BCH/DCI, $C=1$, so this step leaves the sequence unchanged.
\end{enumerate}

At the receiver end, the procedure is as follows:
\begin{enumerate}
\item {\em Code block de-concatenation}.\\
Code block de-concatenation divides the received
LLR sequence into $C$ blocks of lengths $E_r$.
Let $y(0),y(1),\ldots,y(G-1)$ denote the channel
LLRs after demodulation and physical-channel
descrambling, ordered according to the transmitted
coded-bit sequence.
The LLRs associated with the $r$th rate-matched
block are
\begin{equation}
\lambda^{(f)}_{r,k}=y_{S_r+k},
\qquad 0\leq k<E_r,
\end{equation}
where $S_r=\sum_{q=0}^{r-1}E_q$, $S_0=0$,
and $0\leq r<C$.
For BCH/DCI, $C=1$, so no splitting is required.

\item {\em Rate de-matching}.\\
Rate de-matching recovers the length-$N_r$ LLR
sequence required by the polar decoder.
It is performed independently for each code block.

First, coded-bit interleaving is reversed when
$I_{\mathrm{BIL}}=1$.
This operation is omitted for BCH/DCI, for which
$I_{\mathrm{BIL}}=0$.

Next, the bit-selection operation is reversed.
For repetition, LLRs corresponding to repeated
observations of the same coded bit are added.
For puncturing, unobserved positions are assigned
zero LLR.
For shortening, omitted positions are known to
be zero and are assigned a large positive LLR,
using the convention
$\mathscr L=\ln[\Pr(c=0)/\Pr(c=1)]$.
Finally, sub-block interleaving is reversed to
restore the original polar codeword order.

\item {\em Polar decoding}.\\
Polar decoding is performed independently for
each code block.
The decoded input bits are extracted from the
estimated vector $\hat{\mathbf u}_r$, excluding
frozen positions and additional PC positions.
Input interleaving is then reversed when
$I_{\mathrm{IL}}=1$.

Let $h_r(0),h_r(1),\ldots,h_r(K_r-1)$ denote
the resulting estimate of the encoder-input
sequence $c_r(0),c_r(1),\ldots,c_r(K_r-1)$.
Subsection~\ref{subsubsec:polar_dec_algo}
provides more details on polar decoding algorithms.

\item {\em Code block de-segmentation}.\\
For polar-coded UCI, the payload portions of
length $A'/C$ are extracted from the decoded
blocks and concatenated.
The $A'-A$ prepended padding bits are then
removed to obtain
$\hat a(0),\hat a(1),\ldots,\hat a(A-1)$.
The decoded blocks, including their CRC bits,
are retained for CRC verification.

For BCH/DCI, $C=1$ and no code block
de-segmentation is required.
The recovered block still contains its
24 CRC bits.
For DCI, the RNTI-dependent CRC masking
must be reversed before CRC verification.

\item {\em CRC check}.\\
The receiver recalculates the CRC from the
decoded payload and compares it with the
decoded CRC bits.
For UCI, this verification is performed
separately for each decoded segment, including
any prepended padding bits in the CRC input.
For BCH/DCI, the CRC is checked on the single
decoded block using the channel-specific
CRC procedure.
In particular, DCI CRC calculation includes
the prescribed prefix of 24 ones.

The receiver does not have access to the
original transmitted CRC bits.
A mismatch indicates a detected decoding error,
whereas a successful CRC check does not exclude
all possible residual errors.
\end{enumerate}

\subsection{Polar Decoding Algorithms}\label{subsubsec:polar_dec_algo}
Two primary polar decoding methods are SC decoding and BP decoding. Unlike SC decoding, BP decoding does not condition each bit estimate on a sequence of previous hard decisions. It exchanges soft messages between dependent variables on the factor graph. It updates LLR values iteratively through right-to-left and left-to-right passes using message-update functions similar to those used in LDPC decoding. BP decoding supports parallel processing, which can improve throughput, while SC decoding and its variants have sequential decision dependencies that limit, but do not eliminate, parallelization \citep{Ahmadi2018} .
Moreover, polar decoding can employ various algorithms, including SC decoding and SCL decoding.
Polar decoders face challenges in hardware implementation compared to encoders: they process soft information and require recursive computations, message storage, or candidate-path management, depending on the decoding algorithm. These operations can increase latency, hardware usage, and power consumption.
\subsubsection{Successive Cancellation Decoding}~\\
The first polar code decoding method, known as successive cancellation (\acs{SC}), decodes bits one by one, using previous estimates to help determine subsequent bits \citep{Arikan2009}. SC builds on decoded bits sequentially but can suffer from error propagation when an earlier decision is incorrect. Its finite-length performance can be improved by list decoding while retaining its recursive structure. Polar codes achieve the symmetric capacity of binary-input discrete memoryless channels asymptotically under SC decoding. The computation method for SC is akin to that of LDPC codes, using log-likelihood ratios (LLRs) to estimate bit likelihoods.
SC decoding uses LLRs, which can be computed recursively from the synthesized-channel transition probabilities.

Let $W_{N}^{(i)}$ denote the $i$th synthesized bit-channel, and let $y_0^{N-1}$ and $\hat{u}_0^{i-1}$ denote the channel-output sequence and the previously estimated input bits. For $i=0$, the latter sequence is empty. Define
\begin{equation}
\lambda_N^{(i)}\left(y_0^{N-1}, \hat{u}_0^{i-1}\right)=\frac{W_N^{(i)}\left(y_0^{N-1}, \hat{u}_0^{i-1} \mid u_i=0\right)}{W_N^{(i)}\left(y_0^{N-1}, \hat{u}_0^{i-1} \mid u_i=1\right)},
\end{equation}
and let
\begin{equation}
\mathscr L_N^{(i)}\left(y_0^{N-1}, \hat{u}_0^{i-1}\right)=\ln \lambda_N^{(i)}\left(y_0^{N-1}, \hat{u}_0^{i-1}\right).
\end{equation}

This LLR can be used to estimate the value of the bit $u_i$ using the function $h_i$ defined as
\begin{equation}\label{eqn:scl_frozen_bit}
h_i\left(y_0^{N-1}, \hat{u}_0^{i-1}\right)= \begin{cases}0 & \text { if } \mathscr L_N^{(i)}\left(y_0^{N-1}, \hat{u}_0^{i-1}\right) \geq 0 ,\\ 1 & \text { otherwise .}\end{cases}
\end{equation}

Let $\mathcal A$ denote the information set and $\mathcal A^c$ its complement in $\{0,\ldots,N-1\}$. For decoding, if $i\in\mathcal A^c$, then the decoded value is the prescribed frozen bit value, known to the receiver (zero in NR). Otherwise, it is determined by the $h$ function:
\begin{equation}\label{eqn:scl_unfrozen_bit}
\hat{u}_i= \begin{cases}u_i & i \in \mathcal{A}^c,\\ h_i\left(y_0^{N-1}, \hat{u}_0^{i-1}\right) & i \in \mathcal{A} \ .\end{cases}
\end{equation}
Note that $\hat{u}_i$  depends on the previously estimated values $\hat{u}_0^{i-1}$. This is the essence of SC: bits are estimated in order $\hat{u}_0, \hat{u}_1, \hat{u}_2, \ldots, \hat{u}_{N-1}$, with the estimate $\hat{u}_i$ being based upon previously determined bits. The information set is selected to contain reliable synthesized channels, but previous decisions are not guaranteed to be correct~\citep{Arikan2009}. The rule above describes a polar code without additional PC bits. When NR PC bits are present, their values are determined from the preceding decoded bits according to the encoder's parity rule, rather than by an unconstrained hard decision.

\begin{figure}[!ht]
        \centering
        \includegraphics[width=0.5\linewidth]{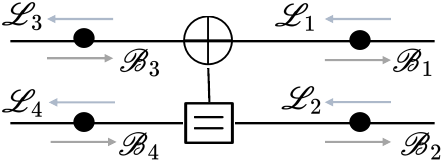}
         \caption{Factor Graph of the SC decoding unit.}
        \label{fig:dec_fact_diagram}
\end{figure}

Moreover, the SC decoding principle can be explained through two elements: the message-update formulas on the unit factor graph and the recursive processing order.
Figure~\ref{fig:dec_fact_diagram} shows the unit factor graph of the polar code, with eight quantities: $\mathscr L_1, \mathscr L_2, \mathscr L_3, \mathscr L_4$ for the LLR values passed to the left and $ \mathscr B_1, \mathscr B_2, \mathscr B_3, \mathscr B_4$ for the binary partial sums passed to the right. At the input-bit positions, these partial sums originate from the bit decisions or prescribed frozen values.

The LLR update equations are
\begin{equation}\label{eqn:llr3_polar}
\mathscr L_3= \mathscr L_1 {\boxplus} \mathscr L_2,
\end{equation}
\begin{equation}\label{eqn:llr_i_polar}
\texttt{\small{where}} \ : \ a \ {\boxplus} \ b =2\tanh^{-1} \left[\tanh \left(\frac{a}{2}\right) \tanh \left(\frac{b}{2}\right)\right].
\end{equation}
\begin{equation}\label{eqn:llr4_polar}
\mathscr L_4= \begin{cases}\mathscr L_2+\mathscr L_1 & \texttt{\small{if }} \mathscr  B_3=0 \ . \\ \mathscr L_2-\mathscr L_1  & \texttt{\small{if }} \mathscr B_3=1 \ .\end{cases}
\end{equation}
\begin{equation}\label{eqn:B1_polar}
\begin{aligned}
   &\mathscr B_1=\mathscr B_3 \oplus \mathscr B_4,\\&\mathscr B_2=\mathscr B_4.
\end{aligned}
\end{equation}

\begin{figure}[!ht]
        \centering
        \includegraphics[width=0.98\linewidth]{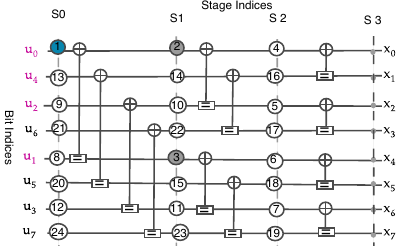}
        \caption{Illustrative instance of a recursive order.}
        \label{fig:dec_recursive_order}
\end{figure}

In what follows, we describe how the recursive process works using the numbered factor graph shown in Figure~\ref{fig:dec_recursive_order}.
First, the decoder needs to find the LLR at point 1. To do this, it needs the LLRs at points 2 and 3. Similarly, the LLR at point 2 depends on those at points 4 and 5, and the LLR at point 3 depends on those at points 6 and 7. The LLRs at points 4, 5, 6, and 7 can be computed directly from the channel LLRs. These upper-branch computations use (\ref{eqn:llr3_polar}). At point 1, the decoder assigns the prescribed frozen value or makes a hard decision according to (\ref{eqn:scl_unfrozen_bit}).

Then, we use (\ref{eqn:llr4_polar}) to calculate the LLR at point 8 from the LLRs at points 2 and 3 and the bit value assigned at point 1. These LLRs are already available, so no further recursion toward the channel is needed. Once point 8 has been decoded, the binary partial sums are propagated to points 2 and 3 using (\ref{eqn:B1_polar}). Their propagation toward points 4, 5, 6, and 7 must wait until the corresponding lower branches have been decoded.

Next, to find the LLR at point 9, we first calculate the LLRs at points 10 and 11. These lower-branch computations use (\ref{eqn:llr4_polar}) and the partial sums at points 2 and 3, respectively. The LLR at point 9 is then obtained from the LLRs at points 10 and 11 using (\ref{eqn:llr3_polar}). Similarly, the LLR at point 12 is computed from those same two LLRs and the bit value assigned at point 9 using (\ref{eqn:llr4_polar}). After point 12 has been decoded, the partial sums at points 10 and 11 are obtained using (\ref{eqn:B1_polar}). The available partial sums at points 2, 3, 10, and 11 can now be propagated to points 4, 5, 6, and 7 using the same XOR relations.
We repeat this process until the SC decoder has assigned all input bits, using prescribed values at frozen positions.
The recursive SC procedure can be implemented using a predetermined node-labelling schedule. The labels in Figure~\ref{fig:dec_recursive_order} specify the processing order for the LLR computations and partial-sum propagation; they do not define a different decoding algorithm.
\subsubsection{Successive Cancellation List Decoding }~\\
The SCL decoder was introduced as an extension of the SC decoder.
Rather than retaining a single hard decision at each information-bit position, it extends each surviving path with both possible bit values. At most $L_{\mathrm{list}}$ paths with the smallest path metrics are retained. Frozen positions do not cause branching, and prescribed PC values, when present, are enforced separately on each path. The path metrics are updated at every bit position, including frozen positions. A list size of 32 can provide performance close to maximum-likelihood decoding for some configurations, but the required list size depends on the code and channel \citep{Ahmadi2018,Tal_vardy2015}.

What is needed is a path metric that ranks the likelihoods of candidate bit sequences. An LLR-based metric is established in \citep{Stimming2014}: for a path $\ell$ with bits $\hat{u}_0(\ell), \hat{u}_1(\ell), \ldots, \hat{u}_i(\ell)$, and for bit index $i \in \{0,1,\ldots,N-1\}$, the path metric is defined as

 \begin{equation}
\mathsf{PM}_{\ell}^{(i)}=\sum_{j=0}^i \ln \left(1+\exp \left[-\left(1-2 \hat{u}_j(\ell)\right) \mathscr L_N^{(j)}[\ell]\right]\right)
\end{equation}
where
\begin{equation}
\mathscr L_N^{(i)}[\ell]=\ln \left(\frac{W_N^{(i)}\left(y_0^{N-1}, \hat{u}_0^{i-1}(\ell) \mid 0\right)}{W_N^{(i)}\left(y_0^{N-1}, \hat{u}_0^{i-1}(\ell) \mid 1\right)}\right)
\end{equation}
is the LLR of the bit $u_i$ given the channel output $y_0^{N-1}$ and the past trajectory of the path $\hat{u}_0^{i-1}(\ell)$. Smaller path metrics correspond to more likely paths. The initialization is $\mathsf{PM}_{\ell}^{(-1)}=0$ for the initial empty path. To avoid overflow when evaluating each increment, use $\ln(1+\exp z)=\max(0,z)+\ln(1+\exp(-|z|))$~\citep{Stimming2014}.

Furthermore, CRC-aided SCL, an extension of the SCL decoder, uses an outer CRC code concatenated with the polar code. This addition helps select a candidate codeword from the final list of paths. When the correct codeword remains in the final list, the CRC can help distinguish it from incorrect candidates. Among the candidates that pass the CRC, the decoder selects the one with the smallest path metric. If none passes, a CRC failure is declared. A successful CRC check does not guarantee that the selected candidate is correct.
Polar decoding via BP and SCAN is beyond the scope of this manuscript. Interested readers are encouraged to refer to \citep{ArliGazi2021, Wang2019} for BP decoding and \citep{Fayyaz2014} for SCAN decoding.
\section{Conclusions}
This work examined the channel coding schemes defined in the 5G NR standard and their associated decoding algorithms, with particular emphasis on LDPC and polar codes. We outlined the design principles underlying these codes, offering insights from both encoding and decoding perspectives. In parallel, we conducted an extensive review of the literature to describe the state of research in this domain.

We complemented these discussions with detailed, standard-specific explanations of procedures that are often difficult to extract directly from technical specification documents. Through this structured approach, the study aims to serve as a valuable reference for researchers and practitioners exploring channel coding in 5G NR and beyond.



\bibliographystyle{unsrt}
\begin{footnotesize}

\end{footnotesize}

\vfill

\end{document}